\documentclass[aps,prb,twocolumn,epsf,epsfig,amsmath]{revtex4-2}
\usepackage{latexsym}
\usepackage{hyperref}
\usepackage{times}
\usepackage{graphicx}
\usepackage{amsmath}
\usepackage{multirow} 
\usepackage{dcolumn}
\usepackage{bm}
\usepackage{ textcomp }
\usepackage{color}
\usepackage{amssymb}
\usepackage{xcolor}
\usepackage{easyReview}
\usepackage{mathdots}
\usepackage{float}
\usepackage{lineno}
\usepackage{todonotes}
\usepackage{array}
\usepackage{amsfonts}   
\usepackage{amssymb}
\newcommand{\beq}{\begin{eqnarray}}
\newcommand{\eeq}{\end{eqnarray}}

\newcommand{\disp}[1]{Eq. (\ref{#1})}

\newcommand{\refdisp}[1]{Ref. [\onlinecite{#1}]}
\newcommand{\figdisp}[1]{Fig. \ref{#1}}

\begin{document}
\title{Incipient quantum spin Hall insulator under strong correlations}
\author{ Peizhi Mai}
\email{peizhimai@gmail.com}
\author{Jinchao Zhao}
\author{Philip W. Phillips}
\email{dimer@illinois.edu}          

\affiliation{Department of Physics and the Anthony J. Leggett institute of Condensed Matter Theory, University of Illinois at Urbana-Champaign, Urbana, IL 61801, USA}

\begin{abstract}
To assess prior mean-field claims that the interacting Kane-Mele model hosts a novel $z-$antiferromagnetic (AFM) Chern insulating phase for a wide range of sub-lattice potentials, we analyze the Kane-Mele-Hubbard model in the presence of a sub-lattice potential using non-perturbative determinant quantum Monte Carlo simulations. We find instead that the true low-temperature state is a quantum spin Hall insulator for intermediate values of the sub-lattice potential $\lambda_v$ and large on-site repulsion.  Two kinds of magnetic fluctuations are found to compete: $z$- and $xy$-AFM.  The latter dominates at low temperature leading to a stabilization of the quantum spin Hall state as opposed to $z-$AFM Chern insulator. Our work is consistent with the robust quantum spin Hall effects  which are consistently observed at even-integer fillings over a wide range of parameters in twisted bilayer MoTe$_2$ and WSe$_2$ as well as AB stacked MoTe$_2$/WSe$_2$.
\end{abstract}
\date{today}


\maketitle



Traditionally, topology and strong correlations lived in different universes.  The former is a function of band structure whereas the latter stems from a breakdown of perturbation theory. These universes now collide with the advent of 2-dimensional moir\'{e} van der Waals materials\cite{XieNature2021,PierceNP2021,ParkNature2021,ChoiNature2021,IpsitaNP2021,NuckollsNature2020,WuNatMat2021,SaitoNP2021,SaitoNP2020,YihangZeng,TingxinLi,JiaqiCai,XuPRX2023,nogapxiaodong,LuNature2024}.  In such materials, strong correlations and topology conspire to yield new phases of matter some of which break time-reversal invariance such as the quantum anomalous Hall (QAH) and ones which preserve it as in the quantum spin Hall (QSH) effect.  A key surprise is that under strong correlations both QSH and QAH can coexist in the same sample\cite{TingxinLi}.  Additionally, zero-field analogues of the fractional Hall effect observed recently in twisted bilayer MoTe$_2$\cite{YihangZeng,JiaqiCai,XuPRX2023,nogapxiaodong} and rhombohedral graphene-hBN moir\'{e} systems\cite{LuNature2024} have further highlighted that interactions and topology examplify ``More is Different''\cite{anderson}. 


Despite these advances, valuable insights can still be gained by studying the standard topological models augmented with interactions.  In particular, twisted MoTe$_2$\cite{FengchengWu,KangNature2024,YihangZeng,JiaqiCai,XuPRX2023,nogapxiaodong} and WSe$_2$\cite{foutty2023mapping} as well as MoTe$_2$/WSe$_2$ heterobilayer\cite{TingxinLi} mimic the Kane-Mele (KM) model under strong correlations. While various studies on the KM-Hubbard model\cite{Hohenadler2011,Hohenadler2012,Rachel1,LessnichPRB2024} consistently reveal a transition from a QSH insulator to a trivial Mott insulator (MI) with $xy$-antiferromagnetism (AFM) at half-filling ($\nu=2$ in experiments) beyond a critical $U_c$, the moir\'{e} transition metal dichalcogenides display QSH effects at even-integer fillings in a range of displacement fields\cite{TingxinLi,foutty2023mapping,KangNature2024}. This suggests that a displacement field may help sustain topology against correlations. Recent {\it mean-field} studies on the KM-Hubbard model\cite{JiangPRL2018,LiuPRL2024}, incorporating a sub-lattice potential $\lambda_v$  (corresponding to the displacement field in experiments) have identified a QAH region with $z$-AFM at half-filling when both $U$ and $\lambda_v$ are large. However, mean-field theory may be useful after the symmetry is known to be broken but is nevertheless prone to exploring symmetry-breaking states in correlated systems. Thus, unbiased methods are essential for investigating the true nature of possible symmetry breaking and emergent topological phases in these correlated systems.


In this study, we solve the KM-Hubbard model with a sub-lattice potential at finite temperatures using the unbiased determinant quantum Monte Carlo (DQMC) method\cite{HuangScience2017,MaiPNAS2024}. We observe that for large $U$ and $\lambda_v$, the system exhibits a QAH-like feature at high temperatures and upon cooling, instead settles into an incipient QSH insulator. Our simulations reveal that $z$-AFM spin correlations are nearly temperature-independent and are generally weaker than the $xy$-AFM spin fluctuations, except in the nearly gapless regime at large $\lambda_v$, where $z$-AFM correlations become only marginally stronger. These results indicate that the QAH state predicted by mean-field theory is an artifact of neglecting strong spin fluctuations. We therefore conclude that the true low-temperature correlated state is a time-reversal-symmetric, incipient QSH phase.
These results are consistent with the ubiquitous QSH effects at even-integer fillings of moir\'{e} transition metal dichalcogenides. 

Calculating the topological invariant in interacting systems is a fundamental and challenging problem. Direct computation of the transverse conductance is difficult. Common approaches include the Niu-Thouless-Wu formula\cite{NTWPRB1985,SinhaPRB2025}, which integrates the Berry curvature over the space of boundary twists and is limited to exact diagonalization, and the $N_3$ invariant\cite{WangPRX2012}.  Recently, we have shown\cite{zhaoprl} that $N_3$ is sensitive to Green function zeros and hence is disconnected from the Hall conductance which can only change if a conducting band crosses the chemical potential.  We therefore adopt neither. Inspired by the experiments\cite{JiaqiCai,foutty2023mapping}, we use the St$\check{\text{r}}$eda formula\cite{Streda_1982,Streda_1983,Bernevigbook}  $\sigma_{xy}=(e/V)(\partial \langle n\rangle/\partial B)_{\mu,T=0}$ ($V$ is the unit cell area), which naturally applies to interacting systems. In an insulator, the Hall conductance is quantized as $\sigma_{xy}=C e^2/h$ and hence the Chern number $C=(1/\Phi_0)(\partial \langle n\rangle /\partial \Phi)_{\mu,T}$ where $\Phi=BV$ is the magnetic flux through each unit cell and $\Phi_0=e/h$ is the magnetic flux quantum. Since the charge gap persists under small magnetic field variations, integrating this formula yields $\langle n\rangle=\langle n\rangle_{\Phi=0}+C(\Phi/\Phi_0)$\cite{Bernevigbook}. 
For QSH effects, when $\hat{S}_z$ is conserved, a generalized St$\check{\text{r}}$eda formula\cite{gstreda,MonacoStreda} obtains for the spin Hall conductance $\sigma_{s,xy}=\sum_\sigma (\partial (\sigma \langle n_\sigma\rangle)/\partial B)_{\mu,T=0}/(2V)$ and the spin Chern number
\beq
C_s=\frac{1}{\Phi_0}\sum_\sigma(\frac{\partial (\sigma \langle n_\sigma\rangle)}{\partial \Phi})_{\mu,T=0}=\frac{1}{\Phi_0}(\frac{\partial \langle n\rangle}{\partial \Phi_{\rm TRI}})_{\mu,T=0}. \label{CS}
\eeq
Here we focus on probing zero-field topology and $\Phi_{\rm TRI}=\Phi\sigma$ represents a time-reversal-invariant (TRI) magnetic flux, inspired by a cold atom proposal\cite{GoldmanPRL2010} to build a spinful TRI Hofstadter system. For insulating states, integrating \disp{CS} similarly gives $\langle n\rangle=\langle n\rangle_{\Phi_{\rm TRI}=0}+C_s(\Phi_{\rm TRI}/\Phi_0)$. To use these algebraic equations, we calculate the compressibility $\chi=\partial \langle n\rangle/\partial \mu$ which vanishes for insulators and is measured experimentally\cite{foutty2023mapping,Jinature2024,WangPRB2023,nogapxiaodong}. Notably, dips in the non-vanishing $\chi(\langle n\rangle)$ at finite temperatures serve as reliable indicators of the $T=0$ insulating states\cite{Haldanequarter,quarter,mfp,DingPRX2024,DingCP2022} (see Supplemental Material\cite{Mai2025SM}). This allows us to infer zero-temperature topology using finite-temperature simulations.

\begin{figure}[t!]
	\centering
	\includegraphics[width=0.5\textwidth]{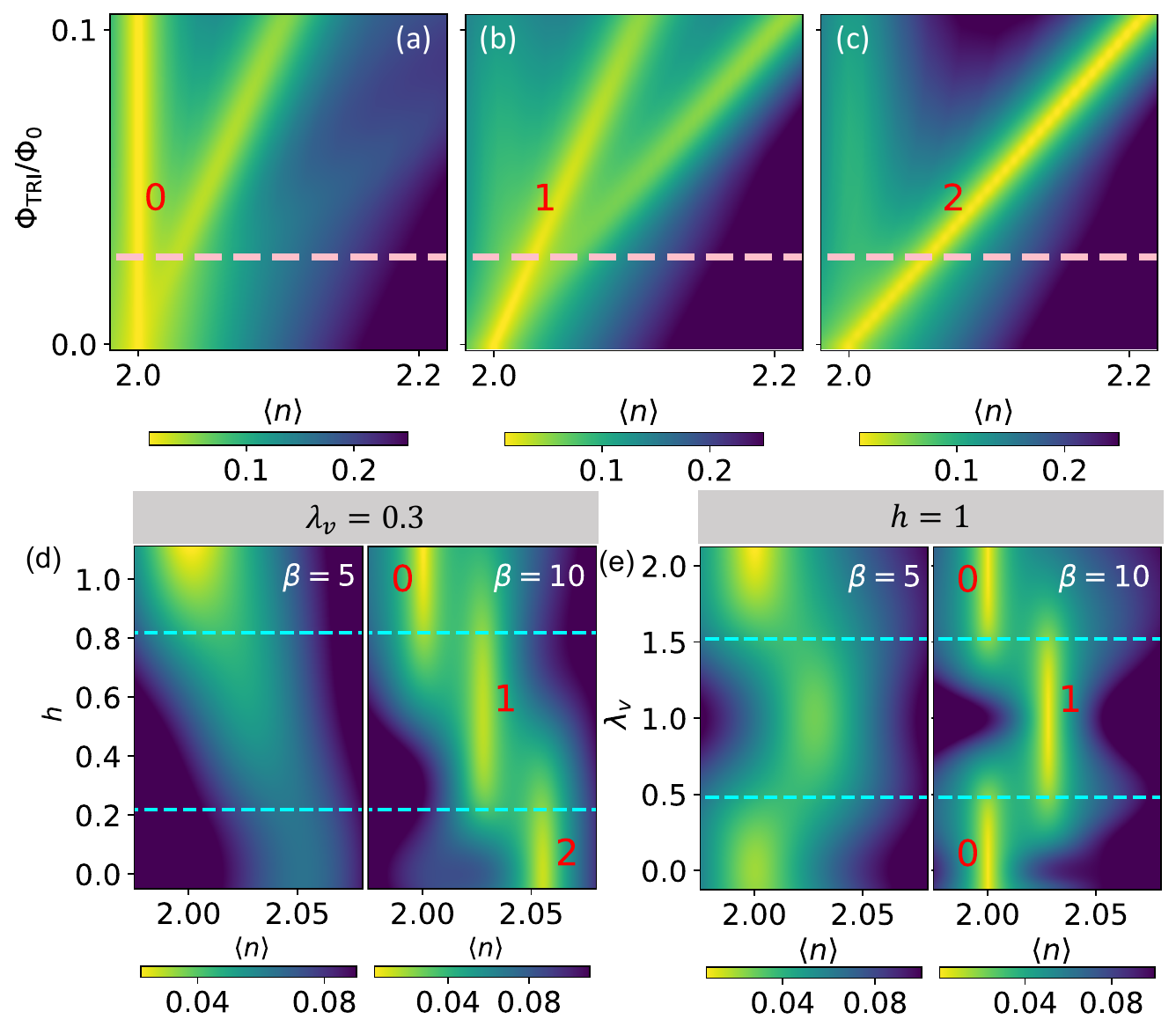}
	\caption{ Panels (a-c) show examples of TRI Compressibility $\chi_{\text{TRI}}(\langle n\rangle, \Phi_{\text{TRI}}/\Phi_0)$ for (a) trivial insulator ($C_s=0$), (b) QAH ($C_s=1$), (c) QSH ($C_s=2$). The inverse slope of the leading valley gives $C_s$ labeled in red. The dashed line gives where we fix the flux $\Phi_{\text{TRI}}/\Phi_0=1/36$ for obtaining panels (d) and (e). Panel (d) show $\chi_{\text{TRI}}(\langle n\rangle,h)$ at the fixed flux with $\lambda_v=0.3$ at $\beta=5$ and $10$. The light regions show the dip and are associated with a $C_s$. Panel (e) show $\chi_{\text{TRI}}(\Phi_{\text{TRI}}/\Phi_0=1/36)$ as a function of $\langle n\rangle$ and $\lambda_v$ with $h=1$ at inverse temperature $\beta=5t^{-1}$ and $10t^{-1}$. The dashed line in panels (d) and (e) depicts the phase boundary. }
	\label{fig:nonintAF}
\end{figure}

We consider the generalized KM model under an external magnetic field,
\beq
\begin{aligned}
\label{Eq:KMBfield}
    H_{\text{KM}}=&-t\sum_{\langle{\bf i}{\bf j}\rangle\sigma} e^{i \phi_{{\bf i},{\bf j}}} 
    c^\dagger_{{\bf i}\sigma}c^{\phantom\dagger}_{{\bf j}\sigma} -\mu\sum_{{\bf i},\sigma} n_{{\bf i}\sigma}\\& -t'\sum_{\langle\langle{\bf i}{\bf j}\rangle\rangle\sigma}e^{\pm i \psi\sigma} e^{i \phi_{{\bf i},{\bf j}} } 
    c^\dagger_{{\bf i}\sigma}c^{\phantom\dagger}_{{\bf j}\sigma},
\end{aligned}
\eeq
where the nearest-neighbor hopping $t=1$ sets the energy scale on the honeycomb lattice. The next-nearest-neighbor hopping $ t^\prime \text{e}^{\pm i\psi\sigma}$ represents the intrinsic spin-orbit coupling via a generalized spin-dependent Haldane phase~\cite{Haldane}, with $\psi = -\pi/2$, unless specified otherwise. To probe the zero field topology using the St$\check{\text{r}}$eda formula and to minimize finite-size effects\cite{mfp} (see Supplemental Material\cite{Mai2025SM}), we introduce an external magnetic field via the Peierls phase $\exp(i \phi_{{\bf i},{\bf j}})$, where $\phi_{{\bf i},{\bf j}}=(2\pi /\Phi_0) \int_{r_{\bf i}}^{r_{\bf j}} {\bf A}\cdot d{\bf l}$ with ${\bf A}=(x\hat{y}-y\hat{x})B/2$. The flux quantization condition $\Phi/\Phi_0=n_f/N_c$ ensures single-valued wavefunctions, where $\Phi=\sqrt{3}Ba^2/2$ is the flux per unit cell, $a$ is lattice constant, $n_f$ is an integer and $N_c$ the number of unit cells. We also consider a TRI magnetic flux $\Phi_{\rm TRI}$ for measuring $C_s$ using \disp{CS}.

We first introduce a symmetry-breaking $z-$AFM mean field ($h\geq0$) and a sub-lattice potential $\lambda_v>0$ to \disp{Eq:KMBfield}. This setup serves both to distinguish different topological phases and to illustrate the underlying mechanism of mean-field theory. The resultant Hamiltonian is
\beq
\begin{aligned}
    H_{\text{KMAFS}}&=H_{\text{KM}}+ \lambda_v(\sum_{{\bf i}\in\text{A},\sigma}-\sum_{{\bf i}\in\text{B},\sigma})n_{{\bf i}\sigma}+h(\sum_{{\bf i}\in\text{A},\sigma}-\sum_{{\bf i}\in\text{B},\sigma})n_{{\bf i}\sigma}\sigma. \label{KMAF} 
\end{aligned}
\eeq
Here we can define an effective spin-dependent sub-lattice potential $\lambda_{v\sigma}=\lambda_v+h\sigma$. We keep $\lambda_v < \lambda_v^{\rm c} = \big|3\sqrt{3}t'\sin{\psi}\big|$, under which the system is a QSH insulator at $h = 0$, and now turn on $h$. 
As $h$ increases within the range $\lambda_{v}^{\rm c}-\lambda_{v}<h<\lambda_{v}^{\rm c}+\lambda_v$, leading to $\lambda_{v\uparrow}>\lambda_{v}^{\rm c}>|\lambda_{v\downarrow}|$, the system transitions into an intermediate QAH phase: spin-down electrons remain in a QAH phase, while spin-up electrons become trivial. As $h$ continues increasing beyond $\lambda_{v}^{\rm c}+\lambda_v$, the topology for both spins becomes trivial. We calculate $C_s$ from the density response to TRI magnetic field using \disp{CS} to distinguish these three phases: QSH with $C_s=2$, QAH with $C_s=1$ and trivial BI with $C_s=0$ . We plot the compressibility as a function of $\Phi_{\text{TRI}}/\Phi_0$ and $\langle n\rangle$ and locate the dominant valley, as illustrated by the light lines of \figdisp{fig:nonintAF}(a-c). For these incompressible states steming from $\langle n\rangle_{\Phi_{\rm TRI}=0}=2$, the algebraic equation is
\beq
\langle n\rangle=2+C_s(\Phi_{\text{TRI}}/\Phi_0). \label{eq4}
\eeq
Hence $C_s$ is given by the inverse slope of the valley, as labeled in red. It is sufficient to fix a small flux (e.g., $\Phi_{\text{TRI}}/\Phi_0=1/36$, as shown by the dashed line in \figdisp{fig:nonintAF}(a-c)) to determine the $C_s$ from the filling. We set $t'=0.1t$, giving $\lambda^{\rm c}_v\approx0.52$. Fixing $\lambda_v=0.3<\lambda^{\rm c}_v$ and gradually increasing $h$, the system can exhibit three different phases (QSH, QAH, BI) as shown in \figdisp{fig:nonintAF}(d). At an inverse temperature $\beta=1/(k_BT)=10$ (in the unit of $t^{-1}$), the valleys appear at densities corresponding to different $C_s$. Next we fix $h=1$, starting from a trivial insulator at $\lambda_v=0$. Increasing $\lambda_v$ past a threshold induces a QAH, with further increases returning the system back to trivial, as shown in \figdisp{fig:nonintAF}(e). While these phases are clearly observed at $\beta=10$, the key features already emerge at $\beta=5$. This example illustrates the underlying mechanism behind the emerging QAH phase in mean-field theory~\cite{JiangPRL2018,LiuPRL2024}: a symmetry-breaking $z$-AFM order induced by strong interactions effectively acts as a spin-dependent potential that combines with the existing sublattice potential. In this picture, an intermediate regime emerges where the combined potential is strong enough to suppress the QAH state of one spin species but not the other, resulting in a net QAH phase for the system.

To examine this picture, we next solve the KM-Hubbard model with a sub-lattice potential:
\beq
\begin{aligned}
    H_{\text{KMHS}}&=H_{\text{KM}}+U\sum_{{\bf i}}(n_{{\bf i}\uparrow}-\frac{1}{2})(n_{{\bf i}\downarrow}-\frac{1}{2})\\&+\lambda_v(\sum_{{\bf i}\in\text{A},\sigma}-\sum_{{\bf i}\in\text{B},\sigma})n_{{\bf i}\sigma}, \label{KMHS}
\end{aligned}
\eeq
using the unbiased DQMC method \cite{mfp,Haldanequarter,quarter,gapopen,DingCP2022,DingPRX2024} on a $6\times6\times2$ cluster restricted by the sign problem (see Supplemental Material\cite{Mai2025SM}). The Jackknife estimate is used to calculate the error bar. The minimal magnetic flux ($\Phi~(\text{or}~\Phi_{\text{TRI}})/\Phi_0=1/N_c=1/36$) is used to accurately determine the zero-field topology using \disp{eq4} and minimize finite-size effects (see Supplemental Material\cite{Mai2025SM} for details). 

\begin{figure}[t!]
	\centering
	\includegraphics[width=0.5\textwidth]{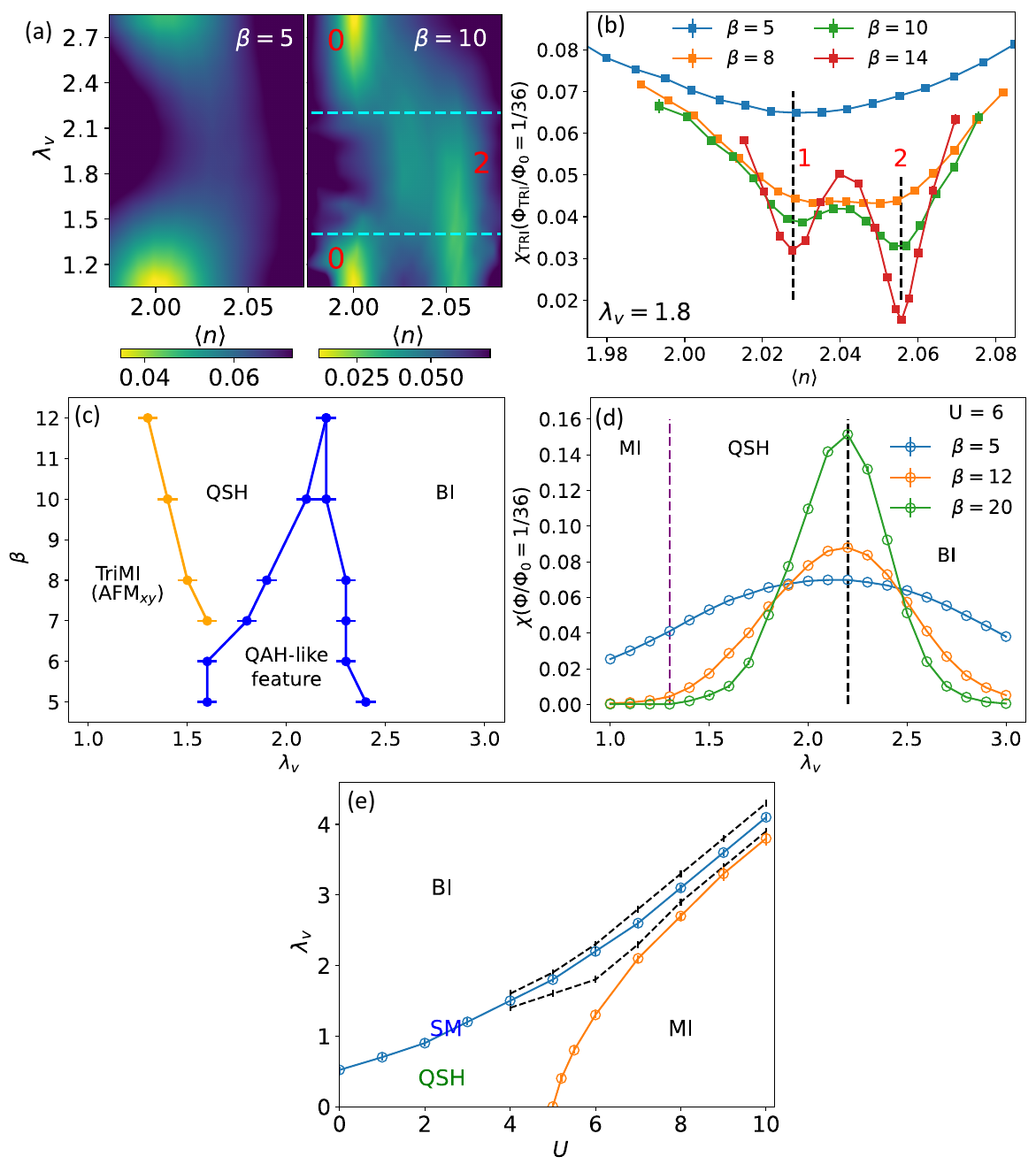}
	\caption{Panel (a) shows $\chi_{\text{TRI}}(\Phi_{\text{TRI}}/\Phi_0=1/36)$ as a function of $\langle n\rangle$ and $\lambda_v$ with $U=6t$ at $\beta=5t^{-1}$ and $10t^{-1}$. The dashed lines label where the topological phase transitions happen. Panel (b) presents the same quantity fixing $\lambda_v=1.8$ and varying temperatures. Panel (c) shows how the topology evolve as $\beta$ and $\lambda_v$ changes. Panel (d) shows the normal compressibility $\chi$ at $\Phi/\Phi_0=1/36$ and $\langle n\rangle=2$. Panel (e) shows the $\lambda_v-U$ phase diagram at the lowest temperatures. Abbreviations: QSH, quantum spin Hall; QAH, quantum anomalous Hall; MI, Mott insulator; BI, Band insulator.}
	\label{fig:U6}
\end{figure}

we continue with $t'=0.1$ which exhibits a sizable QSH gap $\Delta_{\text{QSH}}=2\lambda_v^{\rm c}\approx1.04$ at half-filling when $\lambda_v=U=0$. As $U$ increases, the QSH phase transitions into a trivial MI with ${xy}-$AFM at $U_c=5t$. Fixing $U = 6t$ in the MI regime, we now turn on $\lambda_v$. As indicated by $\chi_{\rm TRI}$ in \figdisp{fig:U6}(a) at $\beta=1/(k_BT)=5t^{-1}$, the system remains trivial for small $\lambda_v$, seems to support $C_s=1$ for intermediate $\lambda_v$ like \figdisp{fig:nonintAF}(e), and becomes a BI for sufficiently large $\lambda_v$. However, upon cooling to $\beta=10$ (\figdisp{fig:U6}(a)), we find a qualitatively different picture. Namely, the leading dip in the intermediate region moves to $\langle n\rangle=2.056$ corresponding to $C_s = 2$, indicating that the true low-temperature state is an incipient QSH insulator rather than a QAH state. This evolution is further clarified in \figdisp{fig:U6}(b), where we track $\chi_{\rm TRI}$ at fixed $\lambda_v = 1.8$. At $\beta = 5$, a $C_s = 1$ dip is present, consistent with QAH-like features. But as $\beta$ increases to 8 and beyond, a second dip corresponding to $C_s = 2$ emerges and eventually dominates, signaling the stabilization of a QSH phase. The full phase evolution as a function of $\lambda_v$ and $\beta$ is summarized in \figdisp{fig:U6}(c). The blue line marks the {\it crossover} between high-temperature QAH-like behavior and either a QSH or BI phase at low temperatures. 

Transitions from the QSH phase to either the MI or BI exhibit behavior distinct from the standard case (\figdisp{fig:nonintAF}(e)), where the valleys of different phases vanish {\it abruptly} at the phase boundary, signaling sharp charge-gap closure. In contrast, \figdisp{fig:U6}(a) shows that at the upper phase boundary, the valleys fade gradually before the transition, indicating an extended gapless region. This is supported by \figdisp{fig:U6}(d), where near transition at $\lambda_v = 2.2$, $\chi$ increases as temperature decreases, suggesting an extended quasi-semimetallic regime. At the lower boundary, the QSH valley persists beyond the transition, while the MI valley dominates at small $\lambda_v$, indicating a transition without closing the charge gap. This behavior extends the earlier findings at $\lambda_v = 0$~\cite{gapopen,Hohenadler2011,Hohenadler2012} to finite $\lambda_v$, demonstrating that such charge-gap-not-closing transitions are a generic class of topological transitions in strongly correlated systems\cite{TingxinLi,EzawaSciRep2013}. To summarize, the phase diagram in \figdisp{fig:U6}(e), based on DQMC simulations at the lowest temperature, reveals an incipient QSH phase at large $U$ and $\lambda_v$, in sharp contrast to the mean-field phase diagrams~\cite{JiangPRL2018,LiuPRL2024}, which predict a $z$-AFM QAH insulator. As $\lambda_v$ increases, the QSH phase transitions into a band insulator (blue line) through an extended quasi-semimetallic regime. Increasing $U$ instead drives a transition into a Mott insulator with $xy$-AFM correlations, but without closing the charge gap (orange line). Both transitions are continuous, one involving charge gap closure and the other accompanied by spontaneous symmetry breaking. These findings highlight not only a qualitatively different phase structure, but also contrasting mechanisms of topological phase transitions compared to mean-field theory.

\begin{figure}[h!]
	\centering
	\includegraphics[width=0.5\textwidth]{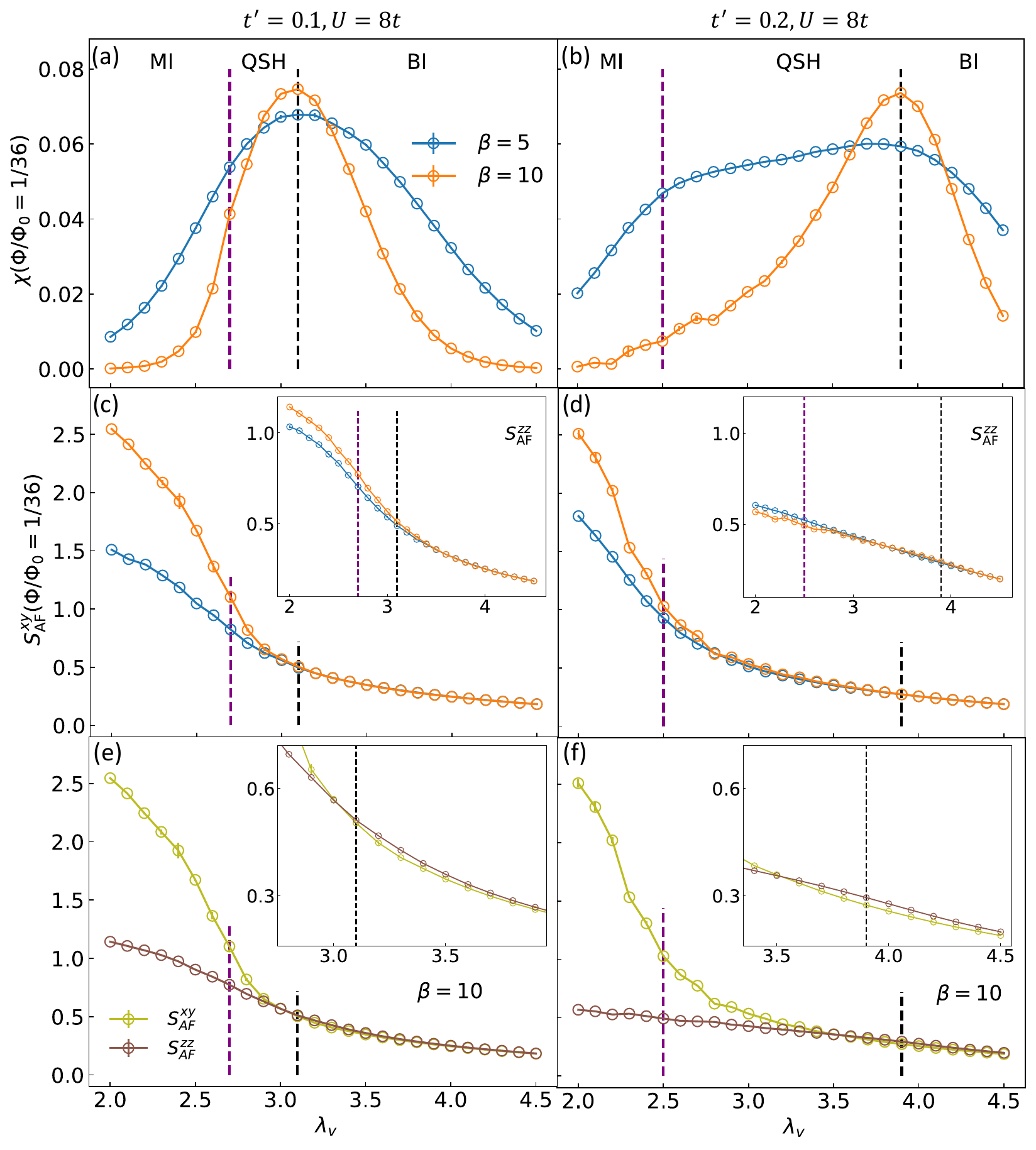}
	\caption{The left and right columns present the normal compressibility (a,b) and AFM correlation (c-f) at minimal flux vs $\lambda_v$ for $t'=0.1$ and $0.2$, respectively, at $\beta=5,10$ and $U=8$. $S_{\rm AF}^{xy}$ is shown in panels (c) and (d), with $S_{\rm AF}^{zz}$ in the inset. Panels (a-d) share the same legend. Panels (e) and (f) compare $S_{\rm AF}^{xy}$ and $S_{\rm AF}^{zz}$ at $\beta=10$.  The dashed lines mark phase boundaries: the left (purple) from a change in $C_s$ via \disp{eq4}, and the right (black) from the compressibility peak.}
	\label{fig:U8}
\end{figure}

The QAH phase proposed by mean-field theory~\cite{JiangPRL2018,LiuPRL2024} relies on spontaneous $z$-AFM order. However, our results point instead to an incipient QSH ground state. To understand this discrepancy, we test the validity of the $z$-AFM assumption by analyzing spin correlations. To ensure the system reaches the possible easy-axis region, we set $U=8t$ and plot the charge compressibility, ${xy}-$AFM ($S^{xy}_{\text{AF}}=(1/N)\sum_{i,j} (-1)^{i+j} \langle(S^x_i S^x_j+S^y_i S^y_j)/2\rangle$) and ${z}-$AFM correlations ($S^{zz}_{\text{AF}}=(1/N)\sum_{i,j} (-1)^{i+j} \langle S^z_i S^z_j\rangle$) correlations under minimal flux as a function of $\lambda_v$ at varying temperatures in \figdisp{fig:U8}. The first column (\figdisp{fig:U8}(a,c,e)) continues using $t'=0.1$. \figdisp{fig:U8}(a) is qualitatively similar to \figdisp{fig:U6}(d) but shows a narrower QSH region due to stronger correlations. As shown in \figdisp{fig:U8}(c), $S^{xy}_{\text{AF}}$ increases along with $\beta$ in the MI phase while $S^{zz}_{\text{AF}}$ is almost temperature-independent in the QSH region. In \figdisp{fig:U8}(e) at $\beta=10$, $S^{xy}_{\text{AF}}$ dominates over $S^{zz}_{\text{AF}}$ for small $\lambda_v$ including the intermediate region. Although $S^{zz}_{\text{AF}}$ slightly exceeds $S^{xy}_{\text{AF}}$ in the quasi semi-metallic and BI region, it does not grow upon cooling, indicating no long-range $z$-AFM order. The second column (\figdisp{fig:U8}(b,d,f)) uses $t'=0.2$, consistent with \refdisp{JiangPRL2018}, with a larger QSH gap $\Delta_{\rm QSH}=2$ when $U=\lambda_v=0$. Hence, a wider QSH region is observed in \figdisp{fig:U8}(b) compared to \figdisp{fig:U8}(a). The $S^{xy}_{\text{AF}}$ correlations are similar between \figdisp{fig:U8}(c) and (d), while $S^{zz}_{\text{AF}}$ is further suppressed for $t'=0.2$. This aligns with the strong coupling analysis\cite{RachelPRB2010}, where the super-exchange from $t'$ and $U$ frustrates $z-$AFM order. In \figdisp{fig:U8}(f), $S^{xy}_{\text{AF}}>S^{zz}_{\text{AF}}$ for most region, except for $\lambda_v>3.6$ where $S^{zz}_{\text{AF}} \gtrapprox S^{xy}_{\text{AF}}$ with little temperature dependence. The case of $t'=0.3,\psi=-\pi/3$ relevant to twisted MoTe$_2$\cite{LiuPRL2024} is similar to the $t'=0.2$ case (see Supplemental Material\cite{Mai2025SM}). Taken together, these results decisively rule out robust $z$-AFM order across all cases studied—including $t'=0.1$, where $S^{zz}_{\text{AF}}$ is relatively enhanced—let alone in systems with larger $t'$ or frustrated $z-$AFM exchange. This directly undermines the key assumption underpinning the mean-field theory~\cite{JiangPRL2018,LiuPRL2024}.

\begin{figure}[h!]
	\centering
	\includegraphics[width=0.5\textwidth]{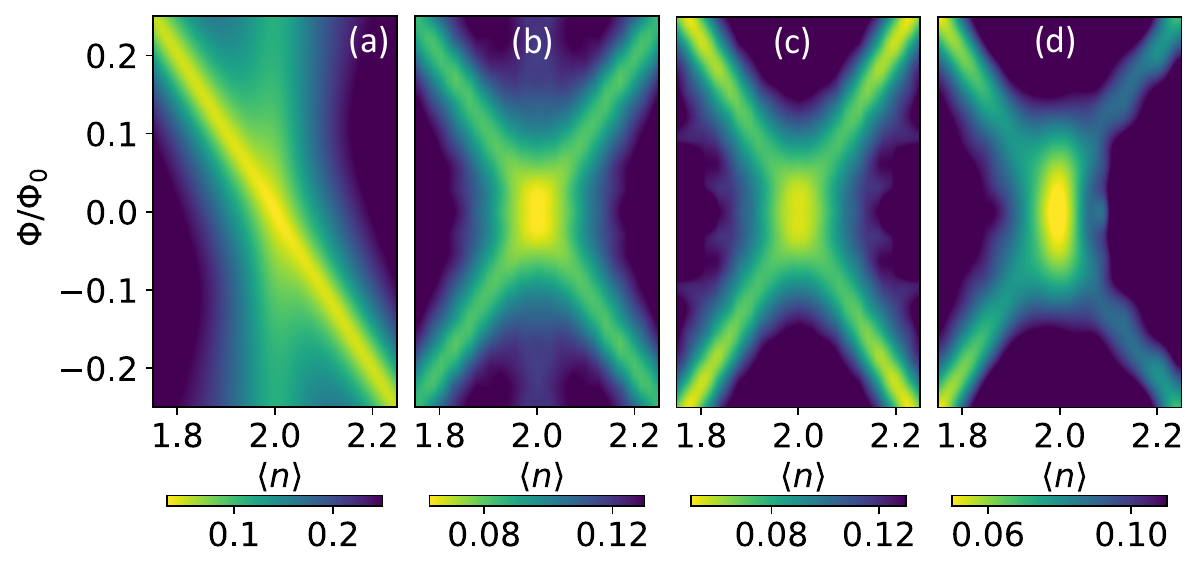}
	\caption{The compressibility $\chi(\langle n\rangle,\Phi/\Phi_0)$ for the non-interacting case (a) $t'=0.1, h=1, \lambda_v=0.6$ and interacting cases: (b) $t'=0.1, U=8, \lambda_v=2.9$,  (c) $t'=0.2, U=8, \lambda_v=3.2$, and (d) $t'=0.3, \psi=-\pi/3, U=8, \lambda_v=3.2$. All panels share $\beta=5$.}
	\label{fig:QSH}
\end{figure}

Further insight into why QSH persists instead of QAH can be gained by examining $\chi(\langle n\rangle,\Phi/\Phi_0)$ for the mean-field (\figdisp{fig:QSH}(a)), and all three interacting cases discussed above (\figdisp{fig:QSH}(b-d)) at $U=8$. A QSH effect displays a short vertical valley at low field indicating $C=0$ and two bifurcating zero Landau levels at high field referring to the splitting of Kramers pair for $\lambda_v\neq0$ (see Supplemental Material\cite{Mai2025SM}), as observed in experiments\cite{KonigScience2008,foutty2023mapping}. In contrast, the QAH state (\figdisp{fig:QSH}(a)) exhibits only one of these branches, with the other suppressed by the combined spin-dependent sub-lattice potential. Instead, the QSH pattern persists in \figdisp{fig:QSH}(b-d) for all interacting cases. Interestingly, the particle-hole symmetry-breaking case simulating twisted MoTe$_2$ in \figdisp{fig:QSH}(d) shows the most robust QSH with highest critical field. That said, while $z$-AFM fluctuations reduce the QSH gap and lower the critical field (most notably in \figdisp{fig:QSH}(b) for $t' = 0.1$), they remain insufficient to stabilize long-range $z$-AFM or induce a QAH phase, in contrast to earlier predictions~\cite{JiangPRL2018,LiuPRL2024}. One might question whether DQMC can in principle host an emergent $z$-AFM Chern insulator at all under strong correlations. This is confirmed in the Haldane-Hubbard model (see Supplemental Material\cite{Mai2025SM}), consistent with earlier unbiased studies\cite{Vanhala,Shao,Imriska,Mertz}. Hence, in the KM-Hubbard model at large $\lambda_v$ and $U$, the true low-temperature state is an incipient QSH state, not the QAH state predicted by mean-field theory.  

We have employed DQMC to study the KM-Hubbard model with a sub-lattice potential $\lambda_v$. We find that the system generally retains a QSH state when both $U$ and $\lambda_v$ are large. This arises because $U$ favors Mott localization across sites, while $\lambda_v$ drives electrons toward one sub-lattice, favoring a band insulator. These competing tendencies partially cancel, allowing the QSH phase to persist in a narrow window as an incipient phase, albeit in a weakened form (with a small gap). At higher temperatures, this regime exhibits QAH-like features, but upon cooling, the system consistently evolves into an incipient QSH insulator, due to the absence of $z$-AFM order suppressed by $xy$-AFM fluctuation. These results directly refute the mean-field prediction of a QAH ground state, establishing instead that the incipient QSH effect emerges robustly from the interplay of strong correlations, topology, and sub-lattice potential. Our study is consistent with the experimental observation that in twisted MoTe$_2$\cite{nogapxiaodong,JiaqiCai,YihangZeng} and WSe$_2$\cite{foutty2023mapping} as well as AB stacked MoTe$_2$/WSe$_2$\cite{TingxinLi,Taozui}, where QSH is consistently observed at even-integer filling. For more quantitative comparison with experiments,
one can fit the tight-binding parameters from density functional theory calculations on the moir\'{e} materials\cite{LiuPRL2024,Devakul,FengchengWu,ZhangNC2024}. For example, $t\approx 10{\rm meV}$ for 3.89$^{\circ}$ twisted MoTe$_2$\cite{LiuPRL2024,ZhangNC2024}, considering $t'/t=0.3$ and $\psi=-\pi/3$ as shown in \figdisp{fig:QSH}(d). Then $\beta=5t^{-1}$ corresponds to $T\approx 2{\rm meV}$ and $U=8t\approx 80{\rm meV}$. 
As $\lambda_v$ increases or $U$ decreases, the QSH state transitions into a BI through an extended quasi-semimetallic region. When $\lambda_v$ decreases or $U$ increases, the QSH state transitions into a MI with ${xy}-$AFM but without charge gap closure. We further demonstrate that such a transition without charge-gap closing is not a fine-tuning exception, but a general class of topological phase transitions in strongly correlated systems.

We thank Xiao Di and Xiaodong Xu for insightful discussions, as well as Cristian Batista for suggesting the pedagogical example of applying an $z$-AFM Zeeman field. This work was supported by the Center for Quantum Sensing and Quantum Materials, a DOE Energy Frontier Research Center, grant DE-SC0021238 (P.M. and P.W.P.). P.W.P. also acknowledges NSF DMR-2111379 for partial funding of the work on QSH in the KM model. P.M. was also supported by the Gordon and Betty Moore Foundation’s EPiQS Initiative through grant GBMF 8691. The DQMC calculation of this work used the Advanced Cyberinfrastructure Coordination Ecosystem: Services \& Support (ACCESS) Expanse supercomputer through the research allocation TG-PHY220042, which is supported by National Science Foundation grant number ACI-1548562\cite{xsede}.

\bibliography{reference}

\end{document}


\title{Incipient quantum spin Hall insulator under strong correlations: supplemental material}
\author{ Peizhi Mai$^{1}$, Jinchao Zhao$^{1}$, Philip W. Phillips$^{1,\dagger}$}

\affiliation{$^1$Department of Physics and Institute of Condensed Matter Theory, University of Illinois at Urbana-Champaign, Urbana, IL 61801, USA}

\date{today}


\maketitle

\section{Compressibility to access insulating states}
The St$\check{\text{r}}$eda formula, introduced in the main text, links topological invariants?namely the charge and spin Chern numbers?to the density response under an external magnetic field. Identifying the filling of insulating states as a function of the magnetic field then becomes the key task. A particularly useful indicator for this purpose is the thermal compressibility,
\beq
\chi=\frac{\partial n}{\partial \mu}=\frac{\beta}{N}\sum_{{\bf i},{\bf j}}\left[ \langle n_{\bf i} n_{\bf j}\rangle - \langle n_{\bf i}\rangle \langle n_{\bf j}\rangle \right],\label{chiformula}
\eeq
which, by definition, vanishes when a charge gap opens. The compressibility can also be computed as the zero-frequency uniform charge correlation function, which is precisely what we calculate in practice during numerical simulations. To demonstrate this, we look at the Kane-Mele (KM) model with a sub-lattice potential $\lambda_v$ under an external magnetic field:

\beq
\begin{aligned}
\label{Eq:KMBfield}
    H_{\text{KMS}}=&t\sum_{\langle{\bf i}{\bf j}\rangle\sigma} e^{i \Phi_{{\bf i},{\bf j}}} 
    c^\dagger_{{\bf i}\sigma}c^{\phantom\dagger}_{{\bf j}\sigma}+t'\sum_{\langle\langle{\bf i}{\bf j}\rangle\rangle\sigma}e^{\pm i \psi\sigma} e^{i \Phi_{{\bf i},{\bf j}} } 
    c^\dagger_{{\bf i}\sigma}c^{\phantom\dagger}_{{\bf j}\sigma} -\mu\sum_{{\bf i},\sigma} n_{{\bf i}\sigma} + \lambda_v(\sum_{{\bf i}\in\text{A},\sigma}-\sum_{{\bf i}\in\text{B},\sigma})n_{{\bf i}\sigma}. \label{KMS}
\end{aligned}
\eeq
We choose the parameters $t=1$ setting the energy scale and $t'/t=0.2, \psi=\pi/2, \lambda_v/t=0.5$. The system is a quantum spin Hall (QSH) insulator at half-filling $\langle n\rangle=2$ and zero field with broken inversion symmetry (due to $\lambda_v\neq 0$) while maintaining particle-hole symmetry.
\begin{figure}[h!]
	\centering
	\includegraphics[width=1\textwidth]{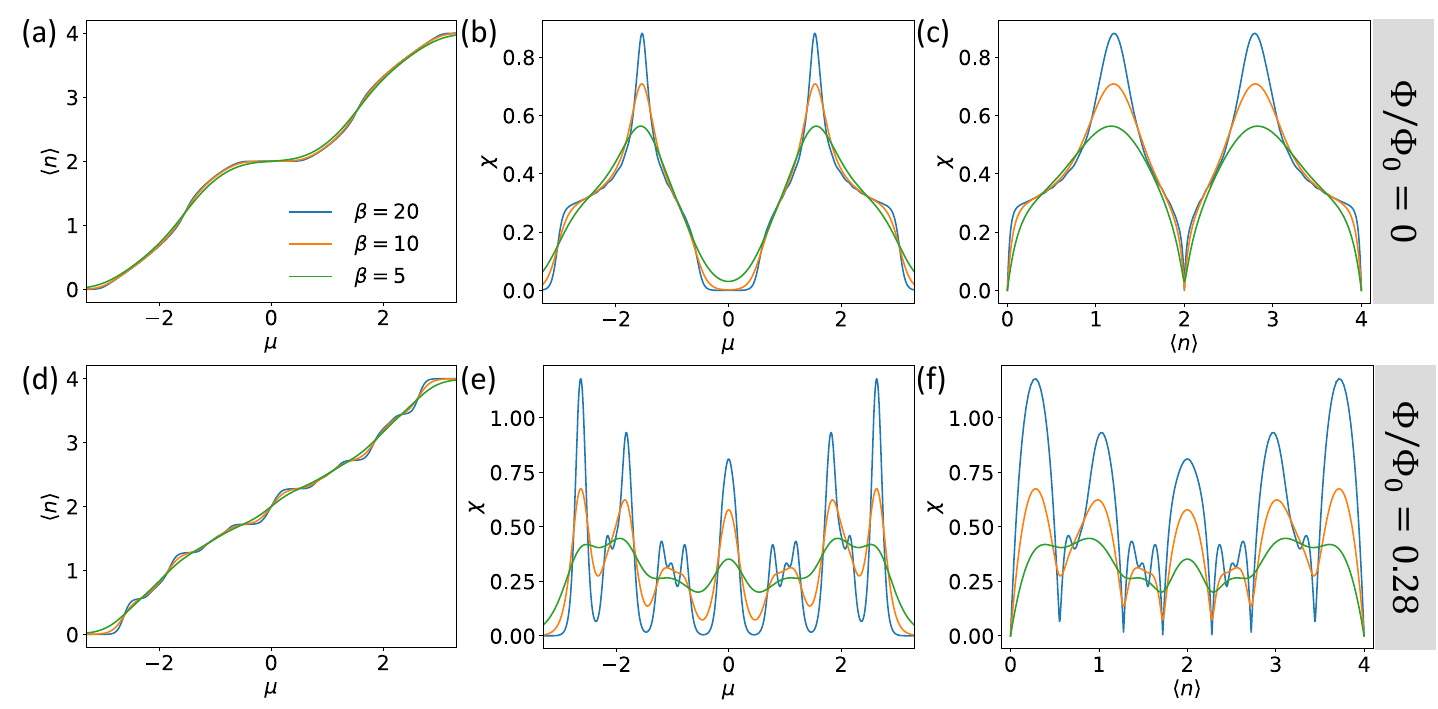}
	\caption{Density $\langle n\rangle$ and compressibility $\chi$ at varying temperatures ($\beta=5,10,20$ in the unit of $t^{-1}$) and magnetic fluxes. $\langle n\rangle$ vs $\mu$, $\chi$ vs $\mu$, and $\chi$ vs $\langle n\rangle$ are shown in first (a,d), second (b,e) and third (c,f) rows, respectively. The first (a-c) and second (d-f) rows correspond to magnetic flux $\Phi/\Phi_0=0,~0.28$ respectively. All panels share the same legend. The shared parameters are $t'=0.2, \psi=\pi/2, \lambda_v=0.5$.}
	\label{supfig:nmuchi}
\end{figure}
When a charge gap opens at low temperatures, an insulating state stabilizes, as shown in the plateaus in \figdisp{supfig:nmuchi}(a, d) for different magnetic fields at $\beta=20$ (in the unit of $t^{-1}$). At finite magnetic field (\figdisp{supfig:nmuchi}(d)), several incompressible (insulating) states emerges at non-zero chemical potential. In \figdisp{supfig:nmuchi}(a, d), as the temperature increases to $\beta=5$, the plateaus soften and become barely visible except for the leading one at $\mu=0$ and zero field. Hence from the plateaus of $\langle n\rangle$ vs $\mu$, it is difficult to find high-temperature precursors of low-temperature insulating states. Now let's look at the compressibility $\chi$ as a function of $\mu$ in \figdisp{supfig:nmuchi}(b,e). When a charge gap opens at low temperature, and $\chi$ vanishes (\disp{chiformula}) as moving $\mu$ inside the gap does not change the density. Even at high temperatures before opening the charge gap, the compressibility has non-vanishing dips indicating precursors of low-temperature insulating states. We need the density information of the insulating states in order to use the St$\check{\text{r}}$eda formula to detect the topology. Hence, we replot the compressibility as a function of $\langle n\rangle$ in \figdisp{supfig:nmuchi}(c,f) given the one-to-one correspondence between $\langle n\rangle$ and $\mu$. Then we know the filling of the insulating states from the dips of compressibility at an external magnetic field even at relatively high temperatures . 

\begin{figure}[h!]
	\centering
	\includegraphics[width=0.9\textwidth]{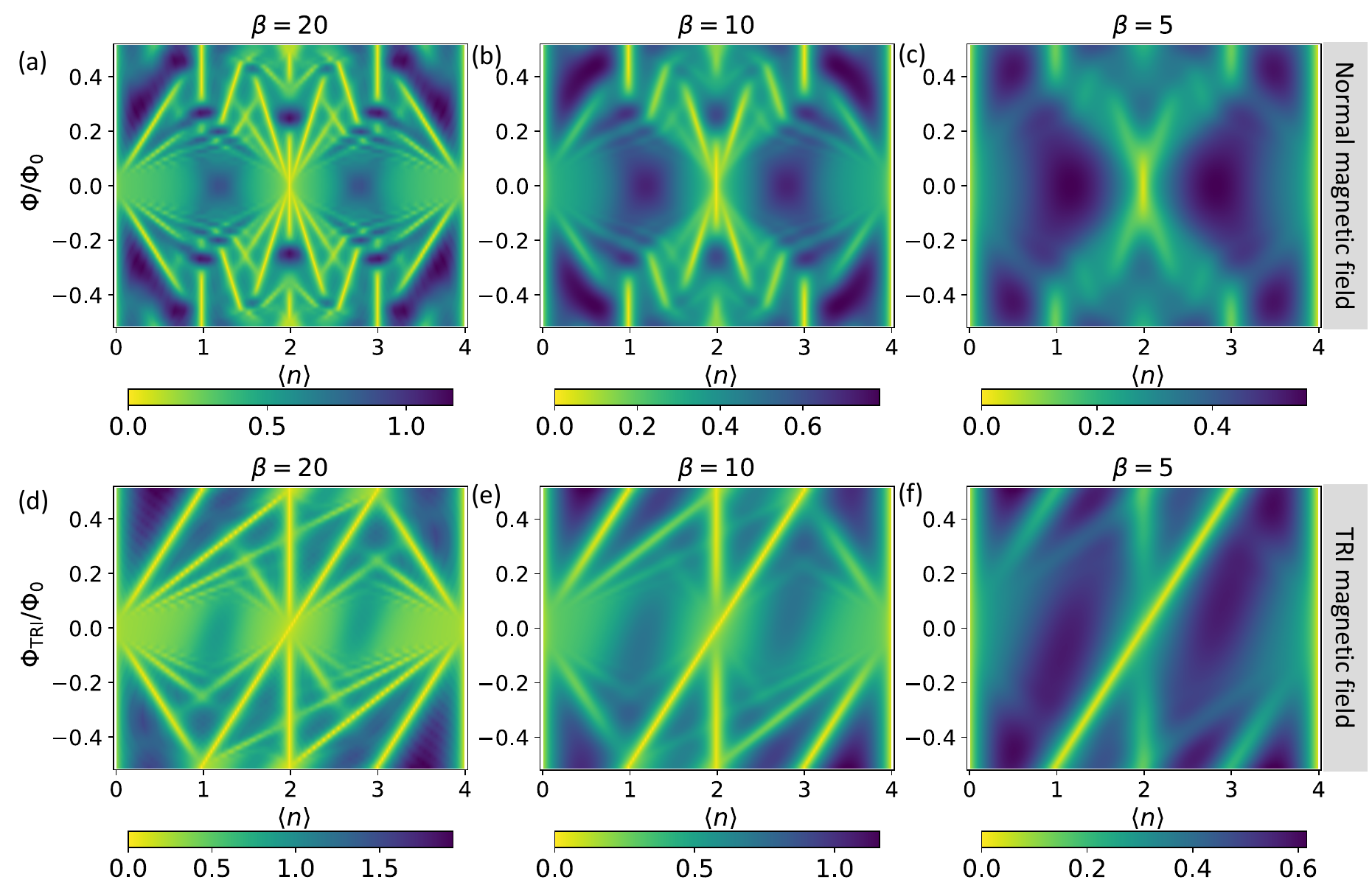}
	\caption{Compressibility $\chi$ as a function of density and magnetic flux (normal (a-c) or time-reversal-invariant field(d-f)) at varying temperatures. The first, second and third columns correspond to $\beta=20$, $10$, and $5$ respectively. The shared parameters are $t'=0.2, \psi=\pi/2, \lambda_v=0.5$.}
	\label{supfig:nonintchivaryT}
\end{figure}

To use the algebraic equation $\langle n\rangle=\langle n\rangle_{\Phi=0}+C(\Phi/\Phi_0)$ derived from the St$\check{\text{r}}$eda formula, we need the $\langle n\rangle$ vs $B$ relation for the incompressible states. Thus, we plot the compressibility in a color plot as a function of $\langle n\rangle$ and magnetic flux $\Phi/\Phi_0$ in \figdisp{supfig:nonintchivaryT}(a-c) at different temperatures. The light region shows the dips in the compressibility, namely the insulating states or their high-temperature precursors. Since we only focus on the zero-field insulating state at $\langle n\rangle=2$, it is sufficient to just look at the high-temperature plot \figdisp{supfig:nonintchivaryT}(c), which already shows the signature of the QSH effect, namely the crossing of two zero Landau levels. From the algebraic equation $\langle n\rangle=\langle n\rangle_{\Phi_{\rm TRI}=0}+C_s(\Phi_{\rm TRI}/\Phi_0)$ obtained from the generalized St$\check{\text{r}}$eda formula for spin Hall conductance, we turn on a time-reversal-invariant (TRI) magnetic field ($\Phi\rightarrow\Phi_{\rm TRI}=\Phi\sigma$) and the corresponding $\chi_{\rm TRI}$ is plotted in \figdisp{supfig:nonintchivaryT}(d) at different temperatures as a function of $\Phi_{\rm TRI}/\Phi_0$ and $\langle n\rangle$. The algebraic equation is only guaranteed to work for probing $C_s$ at zero field because in general $\langle n\rangle_{\Phi_{\rm TRI}=\Phi\sigma,\mu}\neq \langle n\rangle_{\Phi,\mu}$ except for zero field, as one can see from the comparison between the first and second rows in \figdisp{supfig:nonintchivaryT}. Since the purpose is to estimate the $C_s$ at half-filling and zero field, as the inverse slope of the straight-line valley, it is sufficient to look at the high-temperature plot in \figdisp{supfig:nonintchivaryT}(f), which clearly gives $C_s=2$ as expected for the QSH effect.

Therefore, the compressibility is indeed an appropriate quantity to calculate in order to locate the incompressible states under magnetic field and to determine the zero-field topology using the St$\check{\text{r}}$eda formula. In the presence of interactions, it is accurately estimated in the finite-temperature determinant quantum Monte-carlo method by calculating the zero-frequency density-density correlation function.  We obtain this information  without the ill-defined analytic continuation. On the other hand, the compressibility  is often measured in experiments directly using scanning probe microscopy\cite{foutty2023mapping} and indirectly using microwave impedance microscopy \cite{Jinature2024,WangPRB2023} and trion sensing \cite{nogapxiaodong} since the measurement of the compressibility is relatively easier than the direct measurement of the Hall conductance. Thus, it allows a direct comparison between simulations and experimental results. 

\section{Determinant quantum Monte-carlo method}

\begin{figure}[h!]
	\centering
	\includegraphics[width=0.8\textwidth]{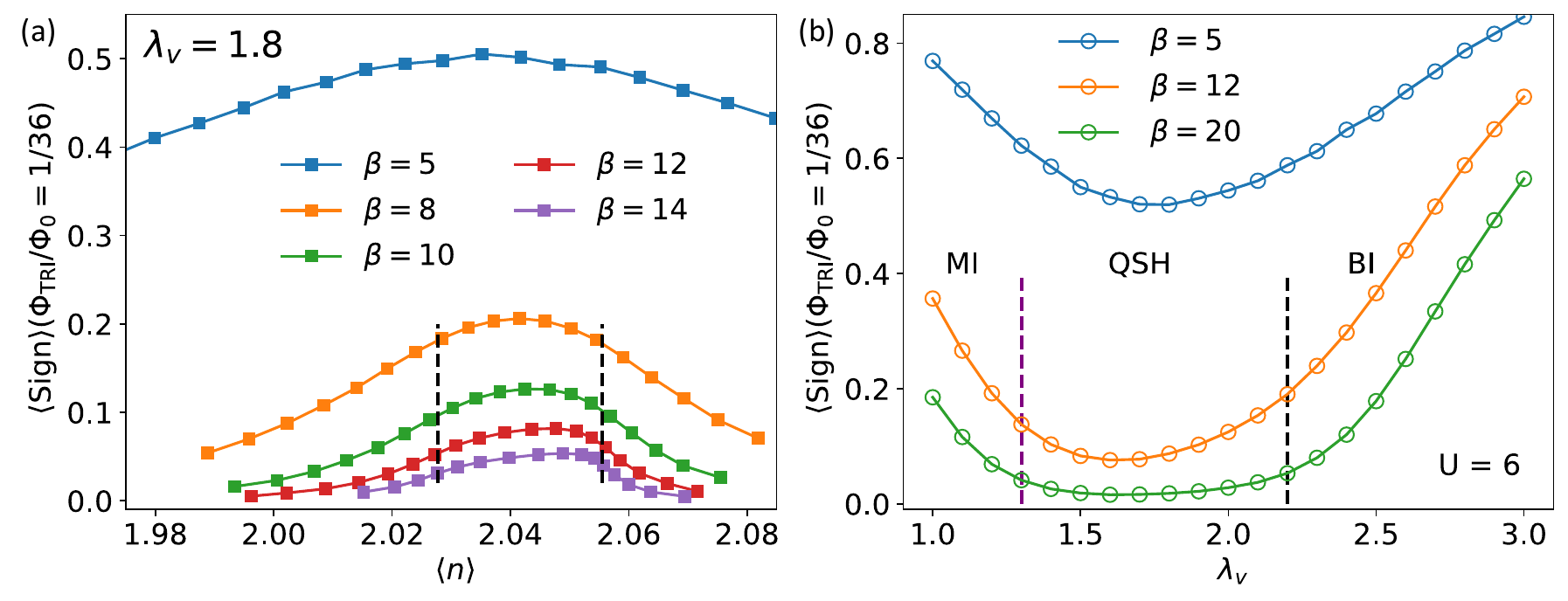}
	\caption{Average sign for KM-Hubbard model with a sub-lattice potential at $U=6$ and varying temperatures. Panel (a) fixes $\lambda_v=1.8$ and shows the average sign as a function of density under a minimal time-reversal-invariant (TRI) magnetic flux. Panel (b) fixes $\langle n \rangle=2$ and presents the average sign as a function of $\lambda_v$ under minimal normal magnetic flux. Both panels share $t'/t=0.1,\psi=-\pi/2$. They correspond to  Fig.~2(b,d) in the main text.}
	\label{supfig:sign}
\end{figure}

The determinant quantum Monte-carlo (DQMC) method\cite{HuangScience2017,MaiPNAS2024} is an unbiased and numerically exact method to solve finite interacting clusters. It have recently been introduced to study interacting topological systems away from half-filling\cite{gapopen,mfp,quarter,Haldanequarter,DingCP2022,DingPRX2024}. We use the DQMC code in \href{https://github.com/edwnh/dqmc}{https://github.com/edwnh/dqmc}. We discretize the imaginary time $\beta$ into $L$ slides with $\Delta\tau=\beta/L=0.1$ and decouple the interaction term by Hubbard-Stratonovich transformation. We then evaluate the partition function through Monte-carlo sampling the configuration of the auxiliary field to obtain the partition function and correlation functions. The KM-Hubbard model with a sub-lattice potential suffers from a sign problem, as illustrated in \figdisp{supfig:sign} corresponding to Fig.~2(b,d) in the main text. 

We restrict to a reasonably large $U=6\sim 10$, low temperature $\beta\sim 10$ and small system size $N_s=6\times6times2$ to avoid an unmanageable sign problem. We conduct $10000$ warmup sweeps and $40000~200000$ measurement sweeps (10 measurements per sweep) at each Markov chain. Depending on the sign problem for the specific parameter set ($U, \beta, \mu$) (little variation under finite magnetic field), we use different numbers (from $2$ to $1000$) of Markov chains to bring down the error bar of the compressibility. 

\newpage
\section{Finite-size effects minimized by minimal magnetic flux}

\begin{figure}[h!]
	\centering
	\includegraphics[width=0.8\textwidth]{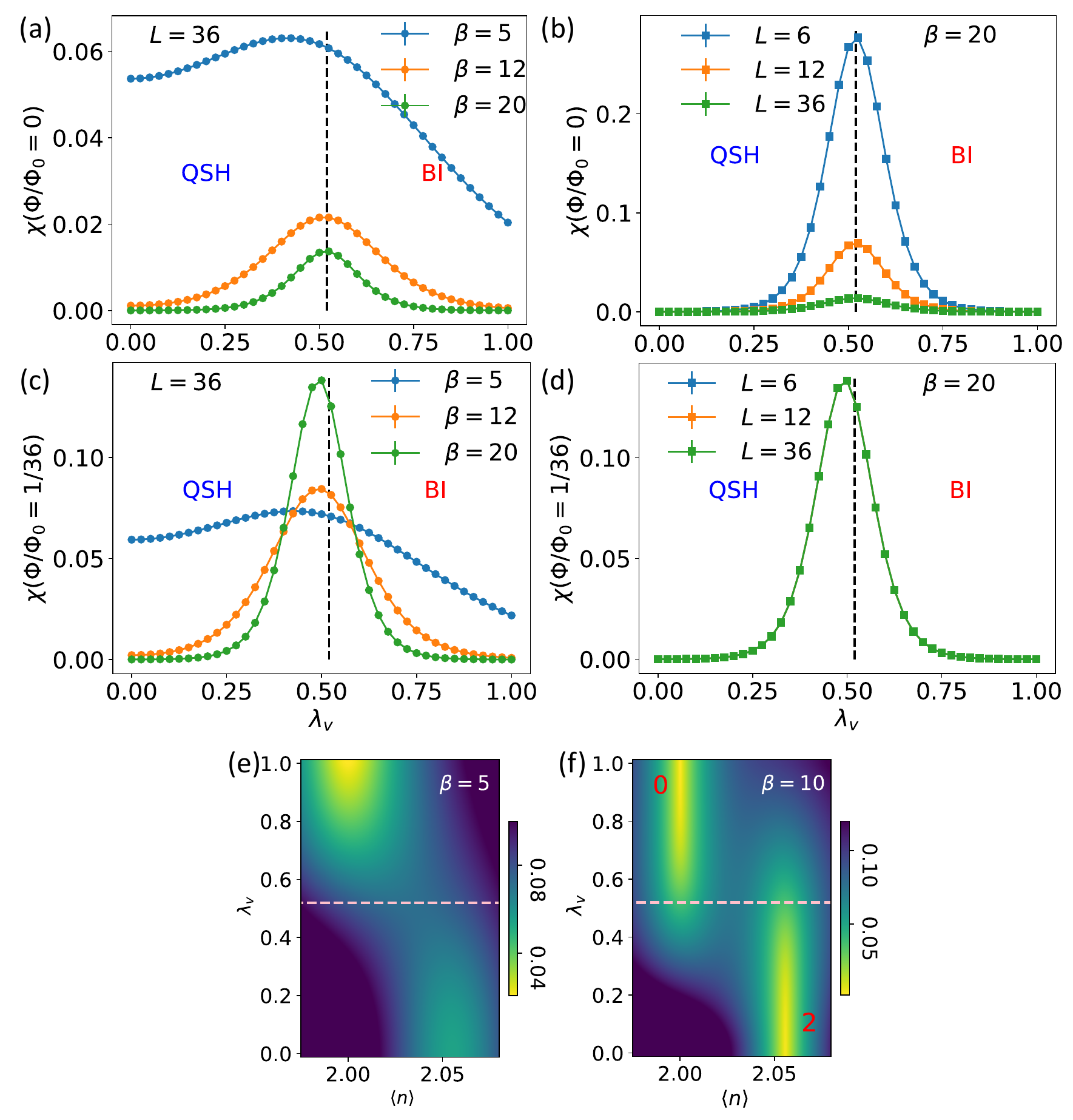}
	\caption{(First row) Non-interacting compressibility at zero flux as a function of sublattice potential difference $\lambda_v$ (a) with varying temperature at $L=36$ and (b) with varying cluster size at $\beta=20$. (Second row) Non-interacting compressibility at magnetic flux ($\Phi/\Phi_0=1/36$) as a function of sublattice potential difference $\lambda_v$ (c) with varying temperature at $L=36$ and (d) with varying cluster size at $\beta=20$. Panels (e) and (f) show the TRI Compressibility $\chi_{\text{TRI}}$ at TRI flux $\Phi_{\text{TRI}}/\Phi_0=1/36$ as a function of $\langle n\rangle$ and $\lambda_v$ with $\beta=5$ and $10$ respectively. The dashed line depicts the transition. All panels share the parameter set $t'=0.1, \psi=-\pi/2, h=0$. }
	\label{supfig:nonintAF0}
\end{figure}

In this section, we consider three topological phase transitions (TPT) mentioned in the main text and show how the finite-size effects can be minimized by turning on a minimal normal or time-reversal-invariant (TRI) magnetic flux. For all these non-interacting example, we employ the flux $\Phi/\Phi_0=1/36=0.028$ or $\Phi_{\rm TRI}/\Phi_0=1/36=0.028$, so that the conclusion applies to the interacting systems where we conduct the determinant quantum Monte-carlo (DQMC) simulations on a $L=6$ cluster (the cluster size is $N_{\rm site}=L\times L\times2$).

In the first example, we look into the Kane-Mele model with a sub-lattice potential $\lambda_v$ under an external magnetic field shown in \disp{KMS}. We fix $t'=0.1,\psi=-\pi/2$ and vary $\lambda_v$. The system is a quantum spin Hall (QSH) insulator for $\lambda_v<3\sqrt{3}t'=0.52$ and a trivial band insulator (BI) for $\lambda_v>3\sqrt{3}t'=0.52$. The single particle charge gap closes at the transition point $\lambda_{vc}=0.52$. We first compute the compressibility without an external magnetic field. The result is shown in \figdisp{supfig:nonintAF0}(a) at varying temperatures $\beta=5,12,20$ for a $L=36$ cluster (assumed to be large enough). The compressibility for all $\lambda_v$ decreases with temperature. The charge gap closes at the TPT, as signalled by the peak of compressibility. Next we gauge the finite-size effect by varying $L$ at the lowest temperature $\beta=20$, as shown in \figdisp{supfig:nonintAF0}(b). The finite-size effect grows as the system approach the phase transition from either side, rendering the results from smaller cluster size unreliable. We then look at the same temperature and cluster size variation respectively in \figdisp{supfig:nonintAF0}(c) and (d) with a minimal magnetic flux $\Phi/\Phi_0=1/36$. Comparing \figdisp{supfig:nonintAF0}(a) and (c), we find that in the presence of the small flux, the compressibility near the transition instead grows with temperature, making the peak more pronounced and thereby facilitating the detection of the transition. That indicates the magnetic flux turns the semi-metal into a metal. We also observe that the location of the peak slightly deviates from the transition by $\Delta\lambda_v=0.025\approx\Phi/\Phi_0$ as a side effect of the magnetic flux. This is acceptable as in the interacting case the $\lambda_v$ interval is $0.1$. Remarkably, in \figdisp{supfig:nonintAF0}(d), all the curves for different system size collapse at the lowest temperature, in contrast to \figdisp{supfig:nonintAF0}(b). There is no visible finite size effect even though we conduct the simulation on the $L=6$. Similar situation also applies to the TRI magnetic flux. Thus, we fix the TRI flux at $\Phi_{\rm TRI}/\Phi_0=1/36$ and plot the TRI compressibility as a function density and $\lambda_v$ at $\beta=5$ and $10$ to observe the phase evolution.

\begin{figure}[h!]
	\centering
	\includegraphics[width=0.8\textwidth]{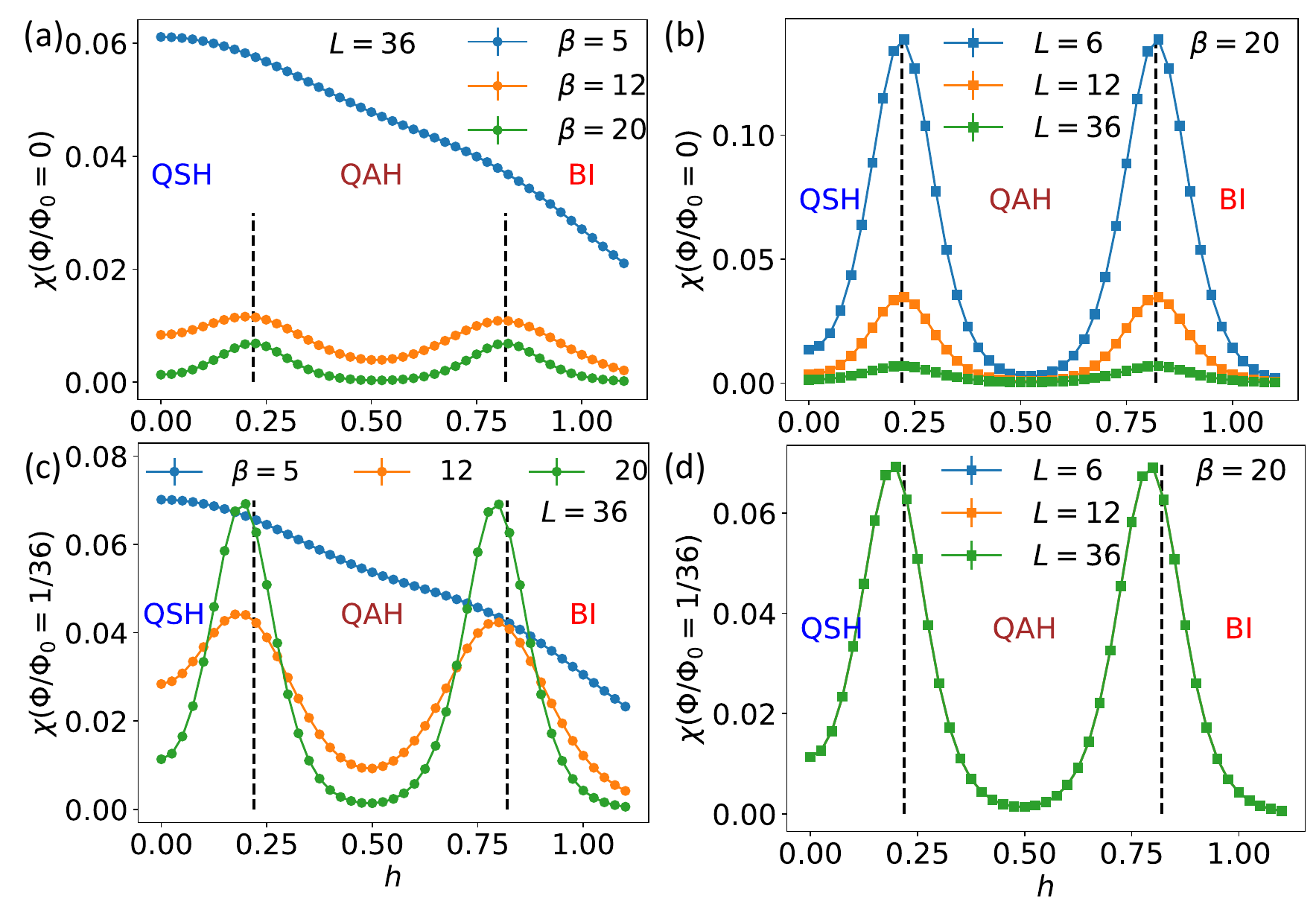}
	\caption{(First row) Non-interacting compressibility at zero flux as a function of AFM$_z$ Zeeman field $h$ (a) with varying temperature at $L=36$ and (b) with varying cluster size at $\beta=20$. (Second row) Non-interacting compressibility at magnetic flux ($\Phi/\Phi_0=1/36$) as a function of AFM$_z$ Zeeman field $h$ (c) with varying temperature at $L=36$ and (d) with varying cluster size at $\beta=20$. All panels share the parameter set $t'=0.1, \psi=-\pi/2, \lambda_v=0.3$.  }
	\label{supfig:nonintlv0p3}
\end{figure}

\begin{figure}[h!]
	\centering
	\includegraphics[width=0.8\textwidth]{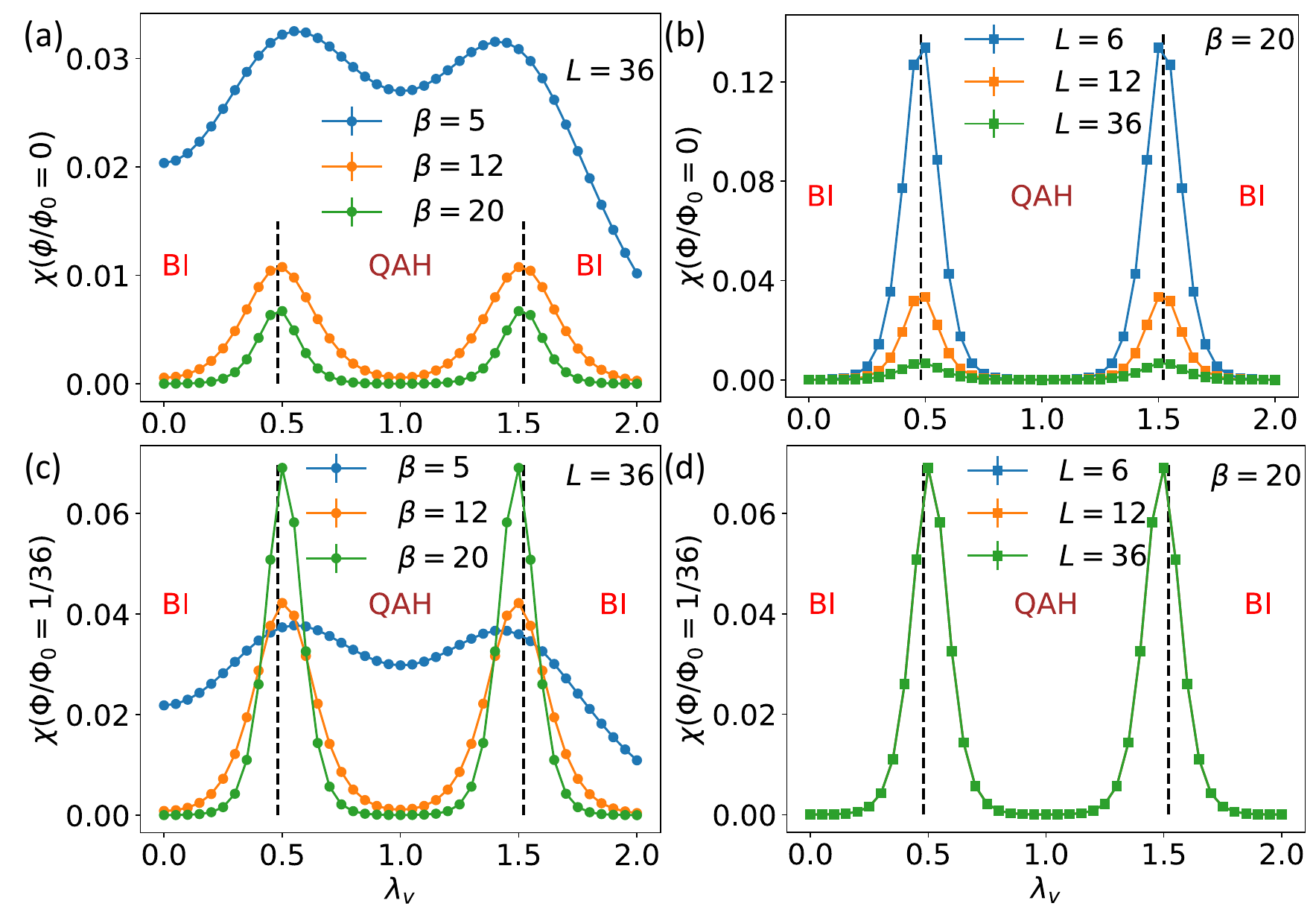}
	\caption{(First row) Non-interacting compressibility at zero flux as a function of sublattice potential difference $\lambda_v$ (a) with varying temperature at $L=36$ and (b) with varying cluster size at $\beta=20$. (Second row) Non-interacting compressibility at magnetic flux ($\Phi/\Phi_0=1/36$) as a function of sublattice potential difference $\lambda_v$ (c) with varying temperature at $L=36$ and (d) with varying cluster size at $\beta=20$. All panels share the parameter set $t'=0.1, \psi=-\pi/2, h=1$.  }
	\label{supfig:nonintAF1}
\end{figure}

Similar behavior is observed in the TPTs when fixing $\lambda_v=0.3$ and increasing the $zz$-antiferromagnetic (AFM$_z$) Zeeman field $h$, as shown in \figdisp{supfig:nonintlv0p3}, and when fixing $h=1$ and increasing $\lambda_v$, as shown in \figdisp{supfig:nonintAF1}. To summarize, employing a small magnetic flux minimizes the finite-size effect and makes the charge-gap-closing transition more pronounced, thought it introduces a small deviation on the estimate of the transition point. 

\clearpage

\section{Landau level feature for QSH effects}
\begin{figure}[h!]
	\centering
	\includegraphics[width=1\textwidth]{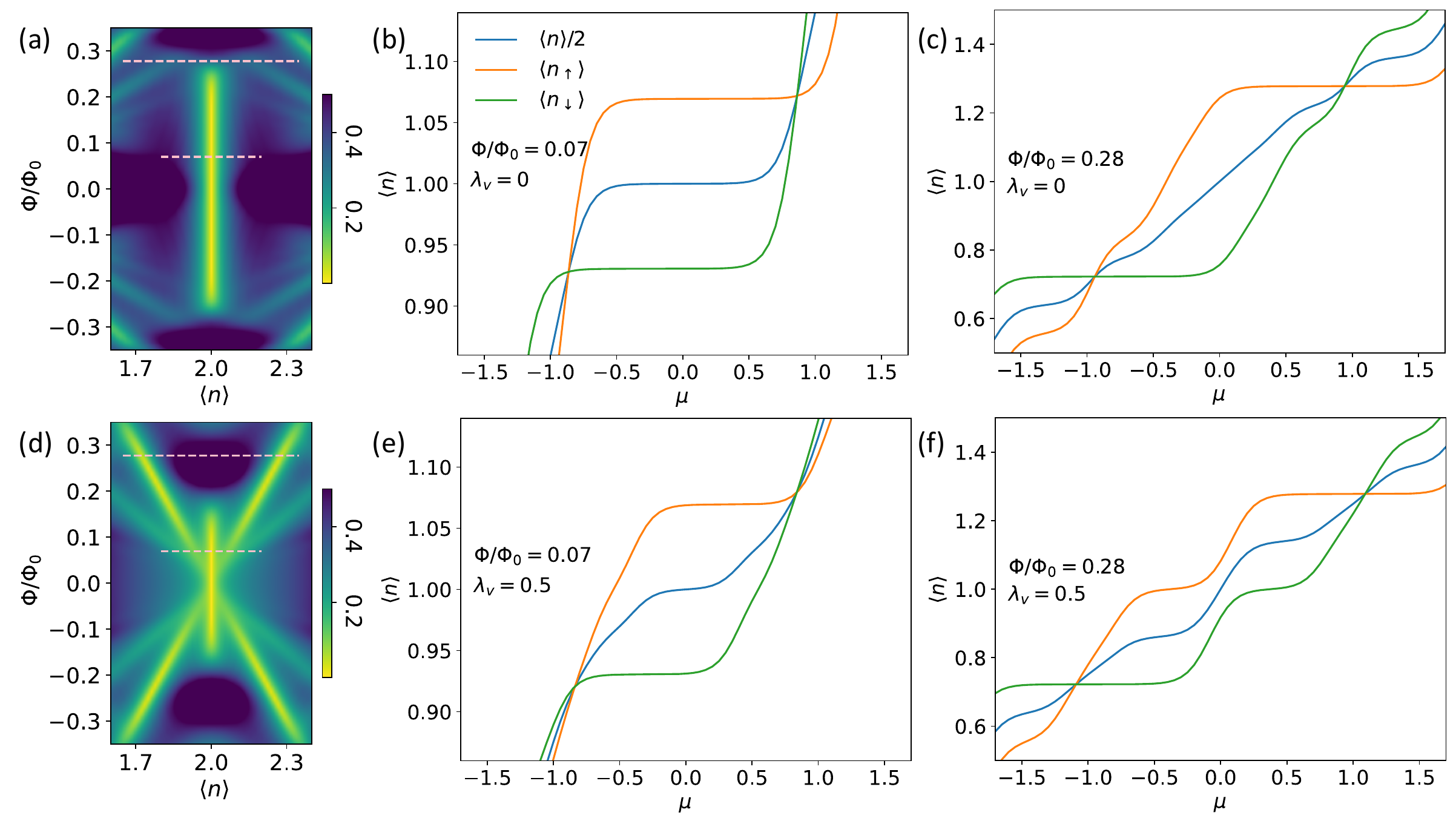}
    \caption{ Panels (a) and (d) show the compressibility as a function of $\langle n\rangle$ and magnetic flux $\Phi$ at $\lambda_v=0$ and $0.5$ respectively with $\beta=12$.  Panels (b) and (e) show $\langle n\rangle$, $\langle n_\uparrow\rangle$, $\langle n_\downarrow\rangle$ all as a function of $\mu$ at fixed $\Phi/\Phi_0=0.07$ for $\lambda_v=0$ and $0.5$ respectively.
    Panels (c) and (f) show the same quantity at fixed $\Phi/\Phi_0=0.28$.  }
	\label{supfig:inversion}
\end{figure}

In this section, we discuss the Landau level signature for the QSH effects. In this discussion we consider the KM model with $t'=0.2,\psi=-\pi/2$. We first set $\lambda_v=0$ and plot the compressibility as a function of magnetic flux and density in \figdisp{supfig:inversion}(a). There is one sharp vertical valley indicating a zero charge Chern number due to time-reversal symmetry. Taking a cut at $\Phi/\Phi_0=0.07$ and $0.28$, we plot the density ($\langle n\rangle, \langle n_\uparrow\rangle, \langle n_\downarrow\rangle$) vs $\mu$ relations in \figdisp{supfig:inversion}(b) and (c) respectively. For small flux $\Phi/\Phi_0=0.07$, the opposite spins carry opposite Chern number and are in incompressible states within an overlapped region of $\mu$, making the combined system an spin Chern insulator. For the high flux $\Phi/\Phi_0=0.28$, there is no overlapped region of $\mu$ where opposite spins are incompressible, there by no valley is observed. Now let's look at the case with $\lambda_v=0.5$, shown in the second row of \figdisp{supfig:inversion}. In \figdisp{supfig:inversion}(d), we observe in addition to the central valley for QSH, two bifurcate Landau levels (LLs) appear at high fluxes signalling the spins carrying opposite Chern number. Taking the cut at the higher flux (\figdisp{supfig:inversion}(f)), we find that while one spin is in a QAH state, the other spin is in a trivial state, thereby making the total system a QAH insulator and explaining the left and right moving LLs. In order to observe such the bifurcate LLs, we need to break the inversion symmetry with $\lambda_v$.

\clearpage
\section{Choice of minimal TRI magnetic flux $\Phi/\Phi_0=1/36$}
In the main text, to estimate the spin Chern number $C_s$ and distinguish the topology of different phases, we introduce a TRI magnetic flux $\Phi_\text{TRI}=\Phi\sigma$, inspired by a cold-atom proposal\cite{GoldmanPRL2010} to build a TRI Hofstadter system. By adding a minus sign to the magnetic flux coupled to spin-down electrons, we preserve time-reversal symmetry even at finite flux. As shown in Fig.~1 of the main text, we estimate $C_s$ from the inverse slope of the leading valley (incompressible state) of the compressibility under TRI flux. Since the valley is a straight line, it suffices to fix a finite value of the flux and infer $C_s$ from the filling of insulating state. This method is particularly useful to locate the transition between QSH and a trivial Mott insulator, which does not involve closing the charge gap and hence leaving no signature on the normal compressibility. 

\begin{figure}[h!]
	\centering
	\includegraphics[width=1\textwidth]{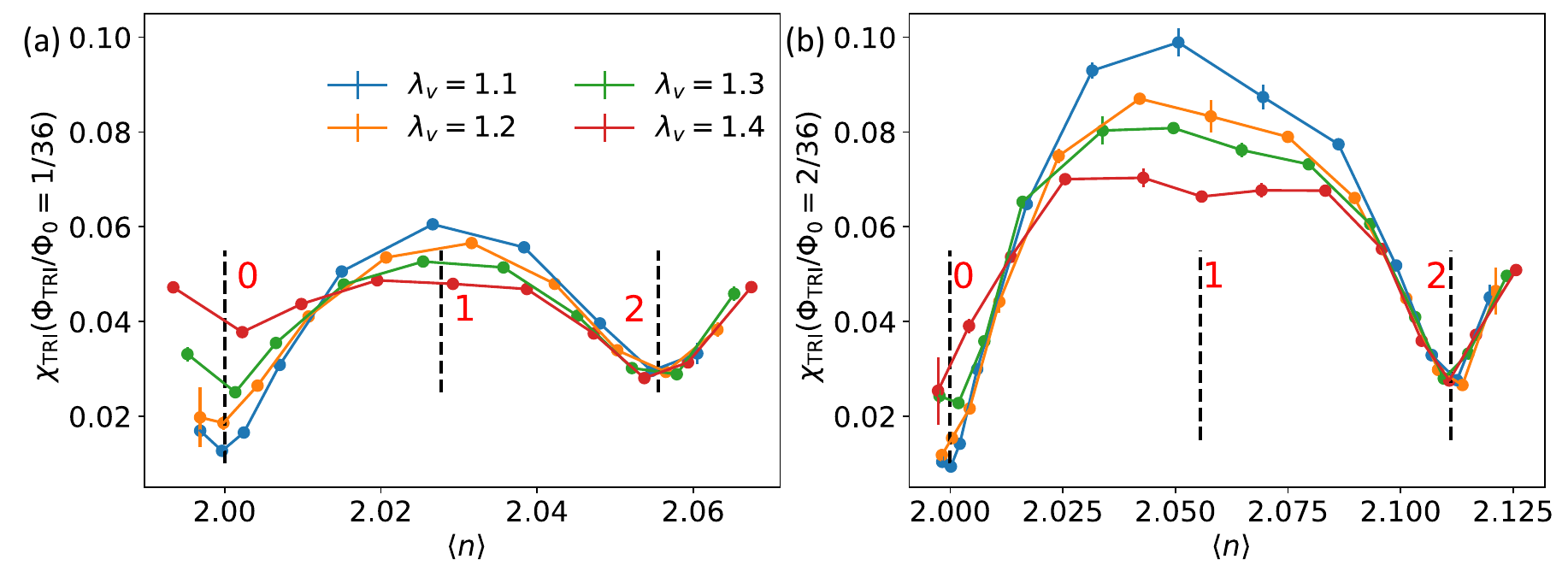}
    \caption{ Panels (a) and (d) show the compressibility at the smallest $(\Phi/\Phi_0=1/36)$ and second smallest $(\Phi/\Phi_0=2/36)$ TRI flux, respectively, in a $6\times6\times2$ cluster as a function of $\langle n\rangle$ at $\beta=10t^{-1}$ for a range of $\lambda_v$.}
	\label{supfig:nflux12}
\end{figure}

For a given topology (QSH, QAH or trivial insulator), the filling of the leading valley up to some finite TRI flux is expected to reflect $C_s$ in the zero-field limit. The safest choice is the minimal flux ($\Phi/\Phi_0=1/36$) for our finite cluster size limited by the sign problem, as illustrated in the TRI compressibility $\chi_\text{TRI}$ in \figdisp{supfig:nflux12}(a). We also present corresponding $\chi_\text{TRI}$ for the second minimal flux $\Phi/\Phi_0=2/36$ in \figdisp{supfig:nflux12}(b) which has similar behaviors, showing the consistency of this approach. However, we observe the change in the dip at $\langle n\rangle=2$ for $\lambda_v=1.4$, which is as low as the dip at $\langle n\rangle=2+C_s*2/36\approx 2.111$. This slightly shifts the phase boundary, and larger TRI fluxes would likely introduce even greater deviations. Therefore, to accurately determine the topology at zero-field limit, we keep the value of the flux pinned to $\Phi/\Phi_0=1/36$. On a technical level, this choice also benefits the DQMC simulation, which appears to have a worsening sign problem around $\langle n\rangle=2$ with larger TRI flux values.

\clearpage

\section{Supplemental data for the phase diagram at $t'=0.1,\psi=-\pi/2$}
\begin{figure}[h!]
	\centering
	\includegraphics[width=0.9\textwidth]{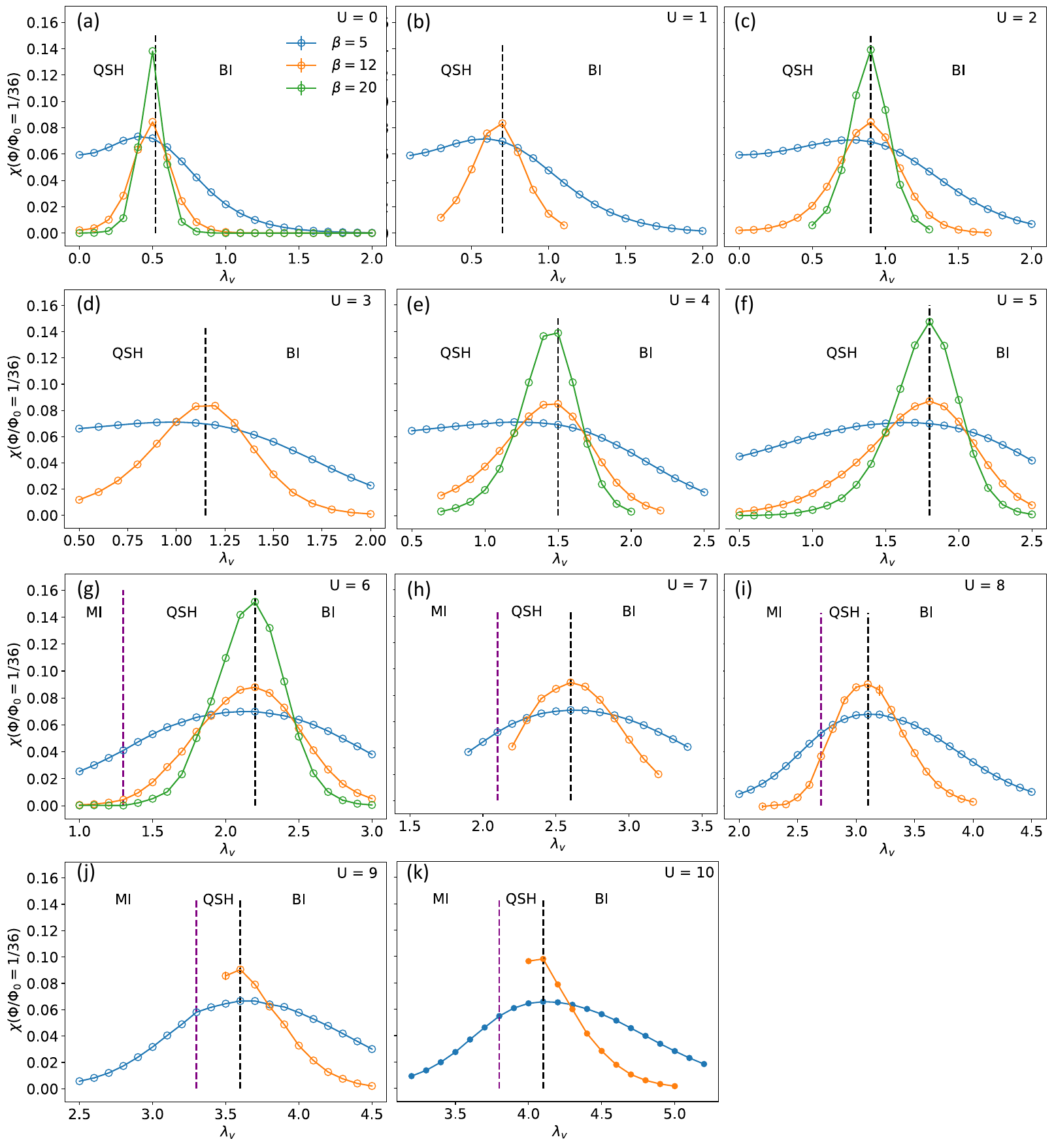}
	\caption{Compressibility at the minimal flux as a function of sublattice potential difference $\lambda_v$ under varying temperature for all $U$ ranging from $0$ to $10$. All panels share the same legend and the parameter set $t'=0.1, \psi=-\pi/2$. The left phase boundary (purple dashed line) is determined from \figdisp{supfig:allUTRI}, while the right phase boundary (black dashed line) denotes the peak of the compressibility. }
	\label{supfig:allUpd}
\end{figure}

\begin{figure}[h!]
	\centering
	\includegraphics[width=0.8\textwidth]{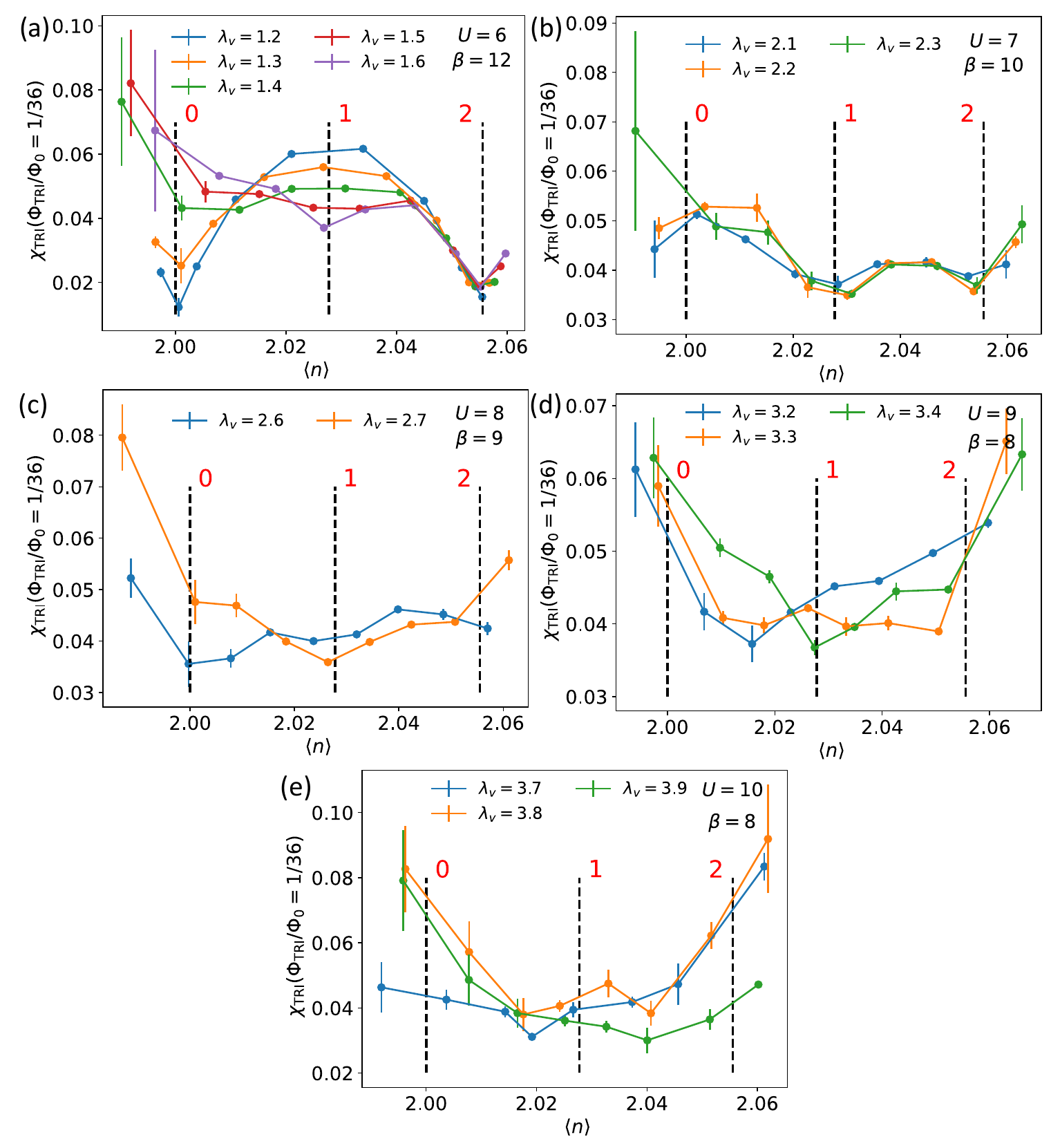}
	\caption{TRI Compressibility at the minimal TRI flux as a function of density for varying $\lambda_v$ at the lowest temperature for all $U$ ranging from $6$ to $10$. All panels share the same parameter set $t'=0.1, \psi=-\pi/2$.  }
	\label{supfig:allUTRI}
\end{figure}

In this section, we provide the complete data set to determine the phase diagram in Fig.~2(e). First we show the compressibility at the minimal flux with varying temperature for all $U$ in \figdisp{supfig:allUpd}. The peak locates the semi-metallic TPT. We observe that as $U$ increases, an extended quasi-semi-metallic region appears around this transition, in contrast to the sharp peak in \figdisp{supfig:allUpd}a at $U=0$. The trivial Mott insulator (TriMI) emerges for $U>5$. To determine the charge-gap-not-closing TPT from QSH to TriMI, we compute the TRI compressibility as shown in \figdisp{supfig:allUTRI} at the minimal flux and lowest temperature restricted by the sign problem. Based on the position of its leading dip, we estimate when the TPT takes place.

\clearpage

\section{Complete data set for compressibility and spin correlations at $U=6$ and $8$}

In this section, we compare the three Kane-Mele parameter sets: $t'=0.1,\psi=\pi/2$ as focused in the main text, $t'=0.2,\psi=\pi/2$ from \cite{JiangPRL2018} and $t'=0.3,\psi=\pi/3$ from \cite{LiuPRL2024} relevant to twisted MoTe$_2$, in the presence of strong correlations $U=6$ and $8$.

\begin{figure}[h!]
	\centering
	\includegraphics[width=\textwidth]{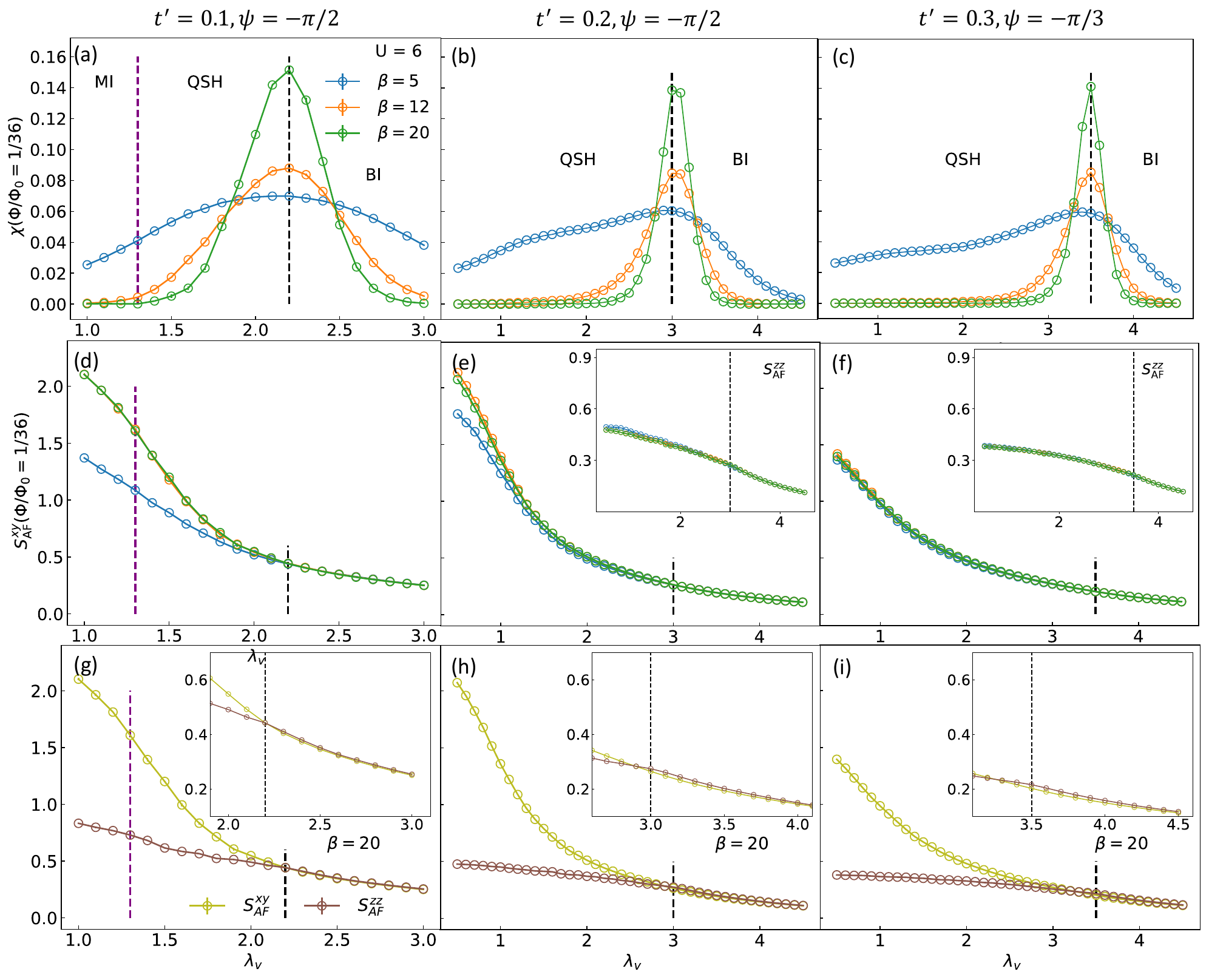}
	\caption{Compressibility and antiferromagnetic (AF) spin correlations at the minimal flux as a function of sublattice potential difference $\lambda_v$ under varying temperature ($\beta=5$ and $12$) for $t'=0.1,\psi=-\pi/2$ (left column), $t'=0.2,\psi=-\pi/2$ (middle column), and $t'=0.3,\psi=-\pi/3$ (right column). The first row shows the compressibility. The second row shows $S^{xy}_{\rm AF}$ and $S^{zz}_{\rm AF}$ (inset) at different temperatures. The third row compares $S^{xy}_{\rm AF}$ and $S^{zz}_{\rm AF}$ at the lowest $T$ ($\beta=10$), with the inset zooming in the region around the semi-metallic transition. All panels share the same legend and are at $U=6$. }
	\label{supfig:U6}
\end{figure}

We present comparison of compressibility and antiferromagnetic (AF) spin correlation among these three cases at $U=6$ in \figdisp{supfig:U6}. $U=6$ is sufficiently strong to obtain a TriMI for $t'=0.1,\psi=-\pi/2$ with a small $\lambda_v$, but not for the other two cases with a stronger original (when $U=\lambda_v=0$) QSH gap. Accompanied with that, the $S^{xy}_{\rm AF}$ is weaker for $t'=0.2,\psi=-\pi/2$ than $t'=0.1,\psi=-\pi/2$, and further weakened for $t'=0.3,\psi=-\pi/3$, as shown in \figdisp{supfig:U6}(d-f). Also, $S^{xy}_{\rm AF}$ increases the most with decreasing $T$ for $t'=0.1,\psi=-\pi/2$, and less for $t'=0.2,\psi=-\pi/2$. It is basically temperature-independent for $t'=0.3,\psi=-\pi/3$, similar to $S^{zz}_{\rm AF}$ in all three cases. The comparison between $S^{xy}_{\rm AF}$ and $S^{zz}_{\rm AF}$ is given in \figdisp{supfig:U6}(g-i). For most of the case, we only observe $S^{xy}_{\rm AF}>S^{zz}_{\rm AF}$, namely an easy-plane. Only in a small region around the TPT for $t'=0.2,\psi=-\pi/2$ and $t'=0.2,\psi=-\pi/2$, shown in the insets of \figdisp{supfig:U6}(h) and (i) respectively, we find $S^{zz}_{\rm AF}$ slightly larger than $S^{xy}_{\rm AF}$. However, as mentioned above, there is little temperature dependence in this region, and the difference is very tiny. Thus, we conclude that there is no easy-axis. 

\begin{figure}[h!]
	\centering
	\includegraphics[width=\textwidth]{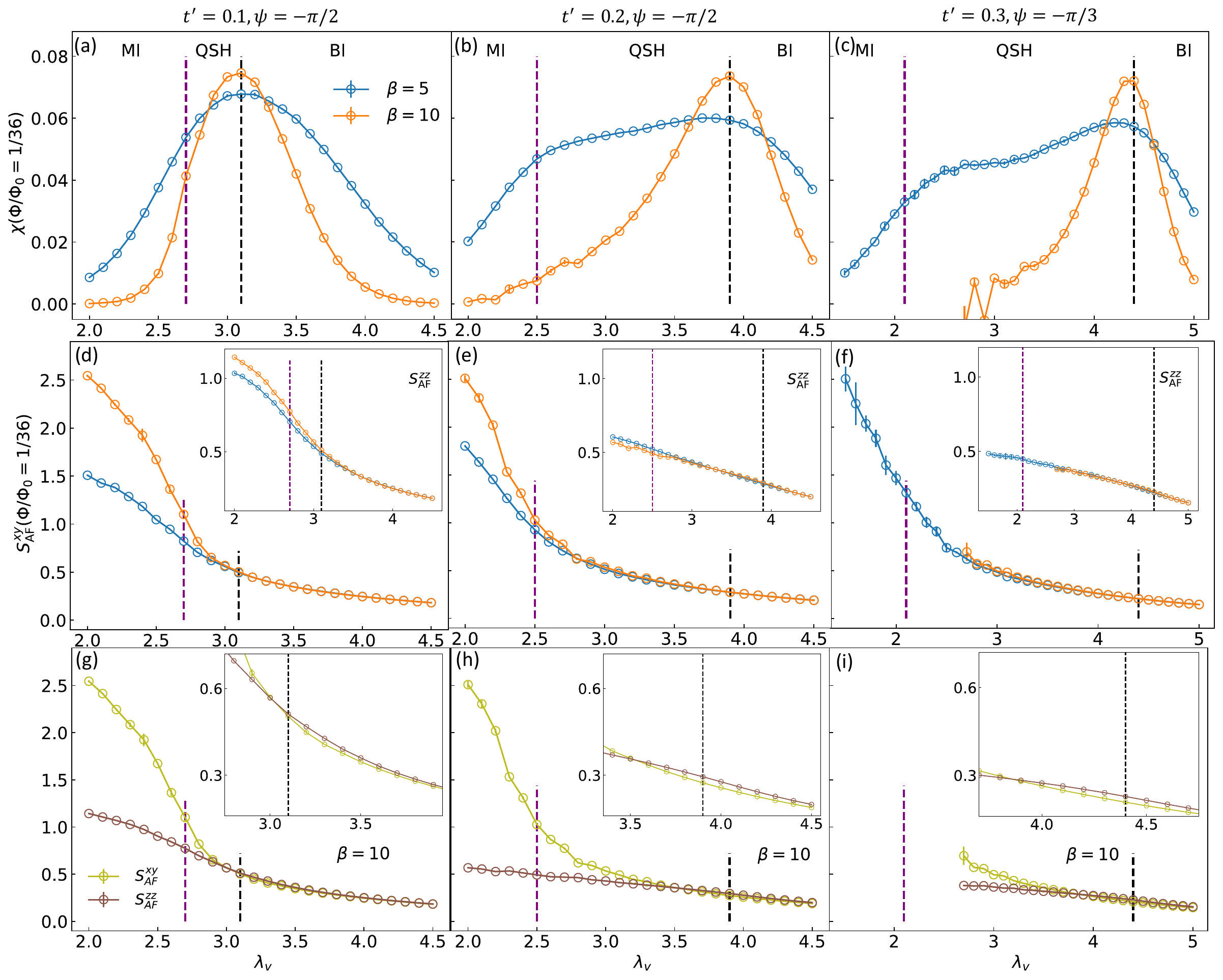}
	\caption{Compressibility and AF spin correlations at the minimal flux as a function of sublattice potential difference $\lambda_v$ under varying temperature ($\beta=5$ and $12$) for $t'=0.1,\psi=-\pi/2$ (left column), $t'=0.2,\psi=-\pi/2$ (middle column), and $t'=0.3,\psi=-\pi/3$ (right column). The first row shows the compressibility. The second row shows $S^{xy}_{\rm AF}$ and $S^{zz}_{\rm AF}$ (inset) at different temperatures. The third row compares $S^{xy}_{\rm AF}$ and $S^{zz}_{\rm AF}$ at the lowest $T$ ($\beta=10$), with the inset zooming in the region around the semi-metallic transition. All panels share the same legend and are at $U=8$. }
	\label{supfig:U8}
\end{figure}

The $U=8$ case is shown in \figdisp{supfig:U8} including the $t'=0.3,\psi=-\pi/3$, compared to Fig.~3 in the main text. Here the $S^{xy}_{AF}$ becomes the strongest in $t'=0.3,\psi=-\pi/3$ at $\beta=5$. However, we can not reach lower temperature due to the sign problem. On the other hand, $S^{zz}_{AF}$ is the most suppressed for $t'=0.3,\psi=-\pi/3$. The comparison between $S^{xy}_{AF}$ and $S^{zz}_{AF}$ is similar to the $U=6$ case. Hence, we again conclude that there is no easy axis.

\clearpage

\section{Antiferromagnetic Chern insulator in the Haldane-Hubbard model}
\begin{figure}[h!]
	\centering
	\includegraphics[width=\textwidth]{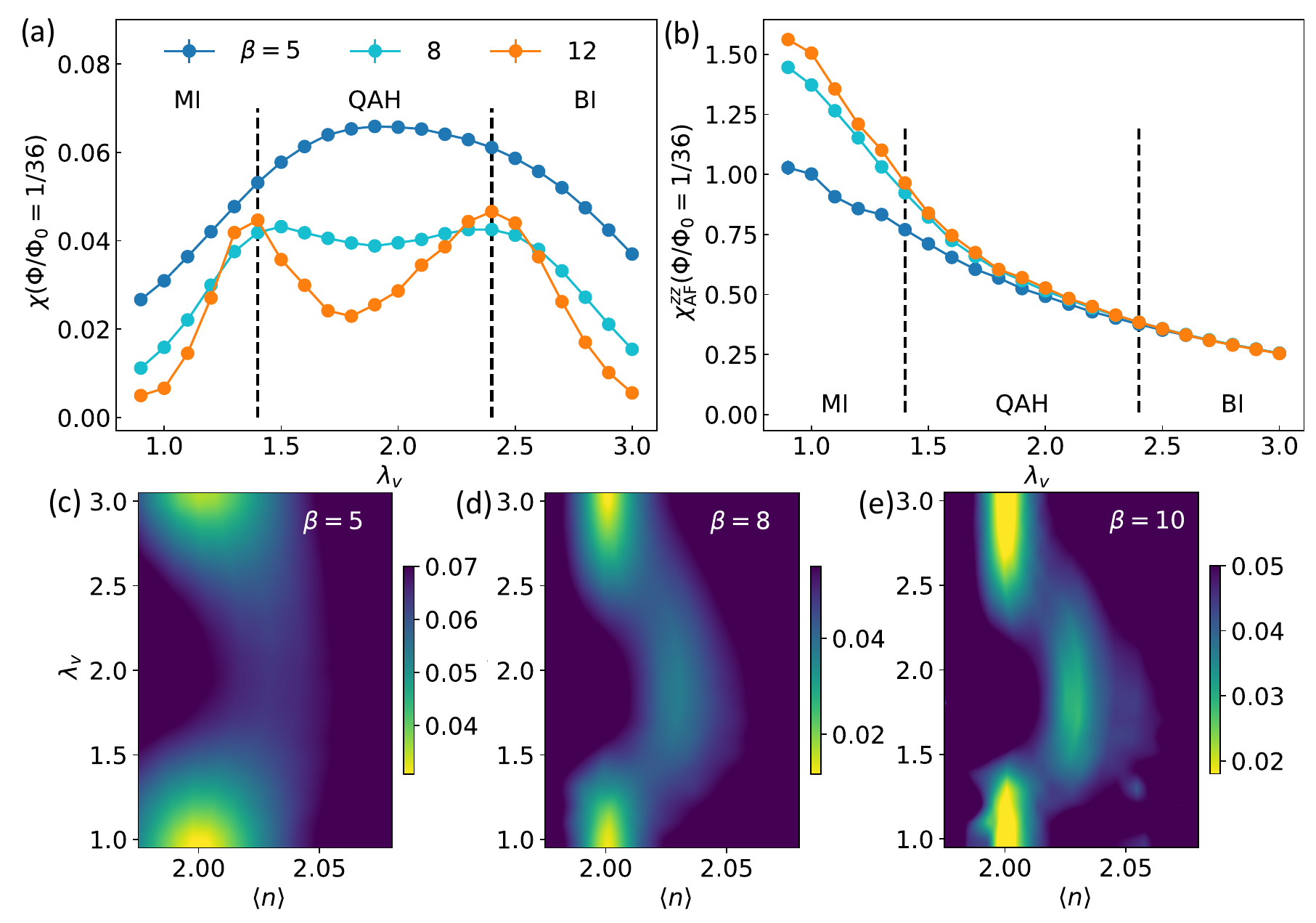}
	\caption{Results from DQMC simulations on Haldane-Hubbard model at $t'=0.1,\psi=-\pi/2,U=6$. The first row shows the compressibility (a) and AF spin correlations (b) at the minimal flux as a function $\lambda_v$ with fixed $\mu=0$. The second row shows the compressibility at the minimal flux as a function $\lambda_v$ and $\langle n\rangle$ at different temperatures. }
	\label{supfig:HaldaneH}
\end{figure}

In this section, we explore a different model, the spinful Haldane-Hubbard model\cite{Vanhala,Shao,Imriska,Mertz}. This model breaks time reversal symmetry explicitly. A quantum anomalous Hall effect obtains at $U=0$ and half-filling with Chern number $C=2$. For large $U$ and $\lambda_v$, an antiferromagnetic Chern insulator (AFCI) obtains with $C=1$, as confirmed by multiple methods\cite{Vanhala,Shao,Imriska,Mertz}. Here we show that our method also supports such an state, in contrast to the Kane-Mele-Hubbard (KMH) case. We show the compressibility at the minimal flux under varying temperatures in \figdisp{supfig:HaldaneH}(a). The double peak structure separate the intermediate topological phase from the trivial states on both sides by a gap-closing TPT. This model maintains SU(2) symmetry. Therefore we only show $S^{zz}_{\rm AF}$ in \figdisp{supfig:HaldaneH}(b) and it grows as temperature decreases. To determine the topology in the intermediate phase, we stick to the minimal flux and plot the compressibility as a function of density and $\lambda_v$ in \figdisp{supfig:HaldaneH}(c-e). As temperature reduces, the Chern number stablizes to $C=1$, consistent with the previous study. These results show our method can spot the AFCI state when it exists and hence support our conclusion that incipient QSH instead of AFCI persists in the KMH model when both $U$ and $\lambda_v$ are large.

\bibliography{reference}